\documentclass[12pt,a4paper]{article}
\usepackage{epsfig}

\usepackage{latexsym}

\newcommand{\be}{\begin{equation}}
\newcommand{\ee}{\end{equation}}
\newcommand{\ba}{\begin{eqnarray}}
\newcommand{\ea}{\end{eqnarray}}

\def\del{\partial}
\def\simge{\mathrel{%
   \rlap{\raise 0.511ex \hbox{$>$}}{\lower 0.511ex \hbox{$\sim$}}}}
\def\simle{\mathrel{
   \rlap{\raise 0.511ex \hbox{$<$}}{\lower 0.511ex \hbox{$\sim$}}}}
\newcommand{\nn}{\nonumber\\ }

\newcommand{\halfs}[1]{\raisebox{-3mm}[0pt][0pt]{#1}}

\def\GeV{{\rm GeV}}                     

\newpage

\begin{document}

\hfill {\bf FIAN/TD-16/03}

\bigskip

\bigskip

\bigskip

\begin{center}
{\bf DENSE GLUON MATTER IN NUCLEAR COLLISIONS}
\end{center}

\medskip

\begin{center}
{\bf Andrei Leonidov}\footnote{E-mail leonidov@td.lpi.ac.ru}
\end{center}

\medskip

\begin{center}
{\it (a) Theoretical Physics Department, P.N. Lebedev Physical
Institute \\ 119991 Leninsky pr. 53, Moscow, Russia}
\end{center}

\begin{center}
{\it (b) Institute of Theoretical and Experimental Physics
\\ 117259 B. Cheremushkinskaya 25, Moscow, Russia}
\end{center}
\bigskip

Theoretical and phenomenological problems of high energy heavy ion
collisions are considered. Main emphasis is on ideas related to
theoretical and phenomenological aspects of Color Glass Condensate
(CGC) physics.

\newpage



\section{Introduction}

Description of strong interaction physics and, in particular, of
multiparticle production processes at high energies in the
framework of Quantum Chromodynamics (QCD) requires a clean
separation of perturbative and nonperturbative elements of the
theory and their relative contribution to the description of the
studied phenomenon. It is well known that a language of elementary
excitations of the theory - quarks and gluons - is literally
applicable only for the description of processes characterized by
large energy-momentum transfer and, correspondingly, by small
spatial and temporal scales. At large times (distances) the theory
finds itself, if viewed in terms of quarks and gluons, in the
strong coupling phase. When moving from the weak coupling regime
towards the strong coupling one, it is necessary to modify the
original perturbative description by adding new elements into the
theoretical picture. Most often this new element is a soft
background field, which encodes a part of nonperturbative
information that is necessary to take into account. New
description amounts to considering  elementary excitations
propagating in this background field. Another possibility is to
directly introduce gluonic strings stretching between constituent
quarks or gluons. Most often a bridge between perturbative and
nonperturbative descriptions is being built  using an idea of
parton-hadron duality, allowing to compare calculations of the
same physical quantities made in partonic and hadronic terms.

The main topic of the present review is theory and phenomenology
of multiparticle production in nuclear collisions analyzed in the
framework of a semiclassical approach developed in the last decade
-- physics of Color Glass Condensate (CGC), see, e.g., the recent
reviews \cite{ILM02,IV03}. The CGC approach puts its main emphasis
on resummation technique within QCD perturbation theory. Other
possibilities for analyzing the high energy nuclear collisions are
simultaneously taking into account both soft (nonperturbative) and
hard (perturbative) degrees of freedom by combining them within
basic Glauber collision geometry \cite{XNW1}, or putting the main
emphasis on the physics of gluon strings stretched between fast
constituent degrees of freedom \cite{KM84,K99}. A detailed
analysis on possible phase transition pattern can be found in
\cite{RFC04}. Multiparticle production in high energy nuclear
collisions is currently a very hot topic - in particular due to
exciting new experimental results from RHIC (see, e.g., the
forthcoming review \cite{E04}).

Let us also stress that many aspects of multiparticle production
are universal, so comparison with the results for $e^+ - e^-$
annihilation \cite{KO97,DG01,D02} or hadron collisions \cite{K03}
is interesting and relevant. What is specific of nuclear
collisions is a very large density of primordial gluonic modes due
to $A$-dependent amplification of the universal perturbative
high-energy (small $x$) growth of the gluon density.

The present review attempts to give a balanced view of both
theoretical (first part) and phenomenological (second part)
aspects of ultrarelativistic heavy ion physics as seen from the
semiclassical CGC-related perspective.

\subsection{Theoretical developments}

To construct a consistent theoretical formalism applicable at high
energies, it is necessary to provide a description for nonlinear
effects for correlators of color excitations in the dense partonic
medium.

One of the classical results of perturbative QCD is a description
of quantum evolution of the correlators of quark and gluon fields
in nuclei with the evolving characteristic scale of the process.
Working in the leading logarithmic approximation (LLA) with
respect to the transferred (transverse) momentum leads to DGLAP
equations \cite{DGLAP}, and in the LLA with respect to the energy
of the process - to BFKL equations \cite{BFKL}. Both DGLAP and
BFKL equations are linear in the parton density.

The necessity of the nonlinear generalization of the linear
evolution equations is particularly clear in the case of LLA in
energy, where logarithmic resummation leads to cross-sections,
fastly growing as a power of energy - a result that does not make
sense. This means that the linear density regime, in which these
equations are valid, contradicts the requirement of unitarity of
the theory and requires generalization. The only cure for this
problem can come from taking into account nonlinear contributions
in parton density. The qualitative importance of such nonlinear
effects and first quantitative calculations were made in the
pioneering work of \cite{GLR83}. Quadratic nonlinearity in parton
density was first considered, in doubly logarithmic approximation,
in \cite{GLR83}, and the coefficient at the nonlinear term was
calculated in Ref. \cite{MQ86}. Phenomenological applications were
discussed in \cite{FS88,LR90}. A lot of attention was devoted to
the description of high-energy QCD asymptotics in terms of
additional reggeon degrees of freedom in the t - channel, see the
review \cite{Li97} and references therein. In particular, in
\cite{Li95} an effective action including both gluonic and reggeon
degrees of freedom was constructed. Let us note, that an
interrelation between the results obtained in the reggeon approach
and in the approach of Wilson renormalization group discussed in
the present review remains at present unclear. This question is a
very important topic for further studies.

A rapid growth of gluonic degrees of freedom in describing the
high energy strong interaction physics lead to an idea of using a
quasiclassical (tree-level) description of the configuration of
color fileds in nuclei as of a basic building block in treating
the high energy processes at some given space-time scale
\cite{MV94}.

A necessity of the collective treatment of gluon modes in the
large density (strong field) regime naturally leads to a picture
of disordered color glass condensate as of a characteristic state
of gluon matter in this limit. The physical picture behind the
color glass condensate is akin to the one considered in the
physics of disordered magnetic systems. More precisely, an average
over the configuration of fast color sources (constituent quarks,
hard gluons) generating the soft gluon field has to be performed
after computing the correlators of quantum gluon modes. It is thus
completely analogous to averaging over disorder in magnetics.
Physically a description in terms of the color glass condensate
arises for fast hadrons (nuclei) propagating with velocity close
to the speed of light. They are observed, because of the Lorentz
contraction, as thin disks propagating along the light cone. The
corresponding formalism is realized as a two-dimensional classical
effective theory valid in some interval of Bjorken $x$
(longitudinal momenta $p^+$). The physical content of the model is
specified by describing the source of the classical gluon field.
The McLerran-Venugopalan model \cite{MV94} model suggests to
consider as such sources the partons (constituent quarks and hard
gluons) carrying a substantial part of the longitudinal momentum
of nucleus. These sources generate a soft gluon field, i.e. gluon
modes with small $x$ in the nuclear wavefunction. The validity of
classical description is related to the large occupation numbers
$\sim 1/\alpha_s$, where $\alpha_s \equiv g^2/4 \pi$ is a strong
coupling constant, so that the corresponding gluon configurations
can be described in terms of strong classical gluon fields ${\cal
A}^i\sim 1/g$.

The McLerran-Venugopalan model is, by construction, a tree-level
classical description of the gluon field generated by constituent
sources in the nucleus. In the low-density (large transverse
momentum) limit its results reproduce the corresponding lowest
order calculations in perturbative QCD. In the limit of high
density of the source (small transverse momentum) the model
predicts {\it saturation} of the gluon distribution. The bending
of the gluon distribution preventing its uncontrollable
"perturbative" growth at small transverse momenta happens at a new
important scale of the theory - saturation momentum $Q_s$. These
features were established in the pioneering calculation of
\cite{JKMW97} and later confirmed in \cite{KM98}.

The considered effective theory is valid for some restricted
interval of $x$. To analyze the contribution of gluon modes with
smaller $x$, it is necessary to move the scale of the effective
theory towards the physical scale of the process $p^+$ by
integrating over the quantum contributions from the gluon modes in
the kinematic interval that "opens" due to the shift of the scale.
Technically the arising procedure is described in terms of Wilson
renormalization group with evolution in rapidity first introduced
in \cite{JKMW97}. For carrying out calculations to all orders in
density requires computing an exact propagator of quantum
fluctuations in the background field. The problem is technically
quite complex and was addressed in a number of publications
\cite{AJMV9596,JKMW97,JKW99,ILM01,FILM02,KM00}.

Historically the major step that had to be made to finalize the
development of the formalism introduced in
\cite{MV94,AJMV9596,JKMW97} was a construction of the correct
effective action, from which it should have been possible to
reproduce, in the linear limit, the BFKL evolution equation. Such
an action was constructed in \cite{JKLW97}, where it was shown,
that the effective action in question contains, in addition to the
usual Yang-Mills term $S_{YM}$, a contribution $S_{W}$, describing
a nonlinear eikonal interaction of the current of fast sources
$J^+$ with {\it quantum} fluctuations of the gauge potential $A^-$
(at tree level $A^-=0$). It is important to note, that the virtual
contribution to the kernel of BFKL equation arises because of
these nonlinear interactions\footnote{One can show, that within
the same technique a "usual" DGLAP equation, corresponding to
evolution in transverse momenta, can also be reproduced
\cite{L00}. For this derivation only the linear contribution from
the eikonal term is needed. The structure of renormalization group
is in this case however more complicated and requires further
analysis. }. Let us also mention Ref. \cite{JJV00}, where the
action analogous to that proposed in \cite{JKLW97} was derived by
considering a physically transparent picture of a system of
precessing color spins and gluon fields.

The next step in understanding the nonlinear effects in QCD at
high energies was made in \cite{JKLW99a}, where a general
functional evolution equation in LLA of the theory was derived.
This equation allows to write a coupled chain of evolution
equations for parton correlators of arbitrary order. It was proven
that for a full description of nonlinear effects in LLA
approximation one should calculate two kernels of the nonlinear
evolution equation, the virtual $\sigma(x_{\perp})$ and real
$\chi(x_{\perp},y_{\perp})$, which are nonlinear functionals of
the background gluon field and generalize corresponding kernels of
the linear BFKL evolution equation. The general nonlinear
evolution equation derived in \cite{JKLW99a} was subsequently
rederived in a number of different ways
\cite{ILM01,FILM02,KW,AM01}. Let us note, that in the limit of
linear kernels the obtained chain of evolution equations formally
coincides with that given by the well known BKP equation, obtained
within the reggeon formalism \cite{BKP80}.

The first explicit calculation of the kernels $\sigma(x_{\perp})$
and $\chi(x_{\perp},y_{\perp})$ was published in \cite{JKW99}. In
this review we will describe in some details the calculation of
\cite{ILM01,FILM02}. Let us note, that the answers obtained in
these papers differ, at least for the light cone gauge and
projectile-centered coordinate system case. The origin of this
discrepancy is currently unclear. Let us note that in the papers
developing the Wilson RG formalism it was shown
\cite{FILM02,KMW00}, that after rotating to covariant gauge and
after coordinate transformation to the target system one
reproduces the equations of \cite{B96,K9900}. This makes the
discrepancy in the answers for the kernels obtained in
\cite{JKW99} and \cite{ILM01,FILM02} very puzzling \footnote{It
looks as if the differences disappear completely after gauge
rotation! }.

In the process of working out the answer for the evolution
equation kernels, in \cite{ILM01,FILM02} a number of results
elucidating the general structure of applying the Wilson
renormalization group formalism to the QCD parton model were
derived. In particular, in \cite{ILM01} a formulation of the QCD
parton model on the complex time contour, called for by the
presence of time-dependent eikonal interactions, was constructed.
This formulation allowed to elucidate a symmetry structure of the
problem and give a rigorous definition of parton correlators as
one-time Wightman functions analogous to that defined in many-body
physics. It was proven, that in the LLA approximation one can use
a real time formulation of the theory.

Results of explicit analytical calculations of
\cite{JKW99,ILM01,FILM02} allowed to compare the evolution of
quantum correlators in Wilson renormalization group with evolution
equations obtained earlier in the framework of operator product
formalsim \cite{B96} and by explicitly calculating Feynman
diagrams in the dipole model formalism \cite{K9900}. The
nonlinearity taken into account in \cite{B96,K9900} effectively
correponds to the triple reggeon interaction term.

An extensive analysis of the solutions of the nonlinear
renormalization group evolution equation has revealed a number of
important and interesting features
\cite{AM99,IM01,IIM02a,IIM02b,LT0001}. The first and, probably,
most important conclusion is that quantum corrections do not
change the basic pattern of nonlinear saturation effects predicted
at tree level by McLerran-Venugopalan model. The quantum evolution
effects show themselves through the energy dependence of
saturation momentum $Q_s$. A second important result described in
\cite{IIM02a,IIM02b} is an effective description of the whole
range of densities (transverse momenta) by a gaussian (mean field)
approximation, in which one has to specify only a two-point
function describing averaging over the sources and providing a
smooth interpolation between the low density and high density
limits.

As mentioned before, a key question addressed by the resummation
program of QCD at high energies is to work out a solution to the
problem of perturbative unitarity violation. Apriori it is, of
course, not clear, whether the purely perturbative solution exists
or whether taking into account effects of all orders in density in
the leading logarithmic approximation through the above-described
nonlinear evolution equation suffices. A detailed analysis of this
problem was recently performed in \cite{FIIM02,KW1,KW2,KW3}. It
turned out that the derived nonlinear evolution equation solves
the problem of unitarity violation only for fixed impact parameter
scattering. It looks quite plausible that the solution of the
problem is, after all, intrinsically nonperturbative. The reason
for this conjecture is a necessity of generating a mass gap in the
spectrum allowing to get an exponential decay of interaction force
corresponding to the exchange of massive paricles. This is
definitely an absolutely nonperturbative phenomenon in non-abelian
teories like QCD.

\subsection{Applications to phenomenology}

The key question of the physics of dense parton medium is a
quantitative understanding of the role of perturbative degrees of
freedom in the early dynamics of nuclear reactions. A clear
example of a formalism, where the hard dynamics can be separated
from the soft one, is a physics of QCD jets, where the hard
primordial parton subprocesses lead to the appearance of
well-collimated fluxes of hadrons in the final state. The
practical aspects of the experimental detection of these fluxes
imposes, however, significant restrictions on the kinematics of
primordial parton scattering (the corresponding minimal transverse
momentum is equal to $50 -- 100\,\GeV$, see e.g.
\cite{LR90,XNW1}). The temptation of generalizing the perturbative
approach to smaller transverse momenta lead to the formulation of
minijet approach to multiparticle production at high energies,
described in the detailed reviews \cite{LR90,XNW1}. The main
(quite drastic) assumption of the minijet physics is a direct link
between the lowest order perturbative diagrams and the inelastic
cross-section. This allows to make an estimate of the {\it number}
of partons that took part in forming the transverse energy flow.
Because of the infrared divergence of the basic $ 2 \to 2$
cross-section the thus calculated perturbative contribution to
inelastic cross section is dominated by the contribution coming
from the vicinity of the infrared cutoff, which has to be
introduced by hand. The estimates made within the minijet
philosophy \cite{JL,BM,KLL,EKL} played a decisive role in the
early estimates of the possibility of producing a dense and hot
partonic matter at the early stages of nuclear collisions.
Technically the estimates of the number of minijet gluons made in
\cite{JL,BM,KLL,EKL} were based on using the lowest order parton
rescattering mechanism and assumption of collinear factorization.
The necessity of considering many binary collisions in the same
event lead to the necessity of using the ad-hoc schemes like
eikonal unitarization \cite{XNW2}.

A special role of minijets in articulating the physical picture of
early parton dynamics describes an interest to the rigorous
analysis of their possible role in the primordial inelasticity
release. Rigorous perturbative calculations are possible only for
the infrared-stable observables (see, e.g., \cite{S}), for which,
if one neglects nonperturbative contributions, the predicted
behavior of the physical observable is fully determined by
perturbation theory. As minijet partonic degrees of freedom can
not be observed as well-collimated fluxes of hadronic transverse
energy, it is natural to consider \cite{LO99} the infrared stable
quantity of transverse energy flow into a fixed rapidity window.
This calculation can be done to the next-to-leading (NLO) accuracy
\cite{LO99}. A detailed analysis of the anatomy of the transverse
energy flow in hadronic collisions \cite{AL00}, made using the
HIJING event generator, shows a dominant role of nonperturbative
degrees of freedom at transverse energies of interest. An
interesting quantity allowing to separate the semihard and soft
contributions to the inelastic cross-section is an azimuthal
asymmetry of the transverse energy flow analyzed in
\cite{LO00,LO01}. The main idea of the cited papers is that the
basic character of semihard and soft mechanism ensures that an
angular asymmetry in the transverse energy flow can arise only due
to the contribution of semihard mechanism, thus allowing to single
out the perturbative contribution.

In the traditional approach to the description of nuclear
scattering, perturbative and nonperturbative components were
considered simultaneously, thus combining the semihard minijet and
soft stringy contributions. Most applications were developed in
the framework of corresponding Monte-Carlo generators HIJING
\cite{HIJING} and PYTHIA \cite{PYTHIA}. One of the most
spectacular results obtained within this approach is a discovery
of a sharply inhomogeneous turbulent nature of the gluon
transverse energy release described in \cite{GRZ97}.

As has been mentioned in the previous paragraph, modern
understanding of the physics of high energy nuclear collisions is
based on the important role of nonlinear interactions in dense
parton medium. The analysis of the role of such effects in
transverse energy production via minijets was first made in
\cite{BM}. The main ingredient of the physical picture of
transverse energy release in high energy nuclear collisions made
in \cite{BM} was that the dominant contribution to the transverse
energy initially produced in these collisions comes from minijets
having transverse momenta of the order of saturation scale. This
hypothesis constitutes the foundation of the saturation
physics-based modern phenomenology of primordial parton dynamics
in heavy ion collisions.

The physical scenario outlined in \cite{BM} was analyzed, from
various points of view, in Refs. \cite{KMW95} - \cite{GM97}. One
of the main directions of research was a development of a
tree-level description of gluon production, generalizing the
formalism of the McLerran-Venugopalan model at calculations of
gluon production. The first complete calculation for gluon
production in the collision of two nuclei in the lowest order in
gluon density was done in \cite{KMW95}. Following the lines of
\cite{KM98}, the problem of finding a gluon spectrum in nuclear
collisions to all orders in density was discussed in \cite{K0102},
where some simplifying assumptions on diagrammatic content of the
answer allowed to obtain an analytically tractable solution. Much
attention was also paid to the numerical analysis of the gluon
production in nuclear scattering \cite{KV99}-\cite{TL03}. This
approach is intrinsically non-perturbative and very promising.

The available data coming from RHIC \cite{PHOB0002}-\cite{STAR01}
made it possible to test the main prediction of Color Glass
Condensate physics in RHIC regime \cite{KN01}-\cite{KLN01},
\cite{BMSS02,M02}. One of the most interesting questions arising
in describing the physics of the early stages of nuclear
collisions is a role of rescattering of initially produced gluons.
In the recent papers \cite{AM00b,BV00,SS01,BMSS02} a detailed
analysis of this question in the framework of saturation CGC
physics, including, in \cite{BMSS02}, a fairly detailed scenario
going beyond binary scattering was made.

The general conclusion is that the experimental data are in
qualitative agreement with the CGC - inspired models, although the
transverse scale characteristic for parton production at RHIC
energies looks somewhat small to consider the usage of
perturbation theory to be reliable.


\section{Tree level description: McLerran - Venugopalan model}

\subsection{Physical picture}

Let us start with describing a physical picture of a heavy nucleus
in the framework of the QCD parton model used in the subsequent
discussion. Let us consider a nucleus moving along the $z$-axis
having four-momentum $P^\mu=(P^0,0,0,P^z)$. In describing
ultrarelativistic particles it is very convenient to introduce the
so-called light-cone coordinates. For some cartesian 4-vector
$v^\mu$ the light-cone coordinates are introduced by the formula
$v^\mu=(v^+,v^-,{\bf v}_\perp)$, where $v^+\equiv (1/\sqrt
2)(v^0+v^3)$, $v^-\equiv (1/\sqrt 2)(v^0-v^3)$, и ${\bf v}_\perp
\equiv (v^1,v^2)$. The scalar product reads $p\cdot x=p^+x^- +
p^-x^+-p_\perp\cdot x_\perp\,$, where $p^-$ и $p^+$ are energy and
longitudinal momentum, and $x^+$ и $x^-$ are light-cone time and
longitudinal coordinate correspondingly.

The physical description of interacting matter within the nucleus
relevant for QCD description divides the parton (quark and gluon)
modes into two basic categories. The first group consists of hard
partons (valence quarks and hard gluons) which carry a significant
part of the longitudinal light-cone momentum of the nucleus $P^+$
and are characterized, in the leading approximation, by the free
motion along the longitudinal $z$ - axis (so that their momenta
are collinear to $P^+$). Hard partons serve as a source for quarks
and gluons with parametrically small longitudinal momenta $q^+\ll
P^+$ -- soft modes.

In the renormalization group approach discussed below it is
crucial to introduce a clear classification of gluon modes in the
light-cone wave function of the nucleus into "soft" and "hard"
ones by comparing their longitudinal momentum $p^+$  with some
characteristic longitudinal scale $\Lambda^+$, so that for the
hard modes $p^+ > \Lambda^+$, and for the soft ones $p^+ <
\Lambda^+$. The scale $\Lambda^+=x_0 P^+$ should be such that
$x_0$ should not be too small. Physically this corresponds to the
condition $x_0 \gg x$, where Bjorken $x$ characterizes the
longitudinal scale of the probe interacting with the nucleus. In
what follows we will be interested in the small-$x$ domain $x\ll
1$. In this regime the light-cone wave function of the projectile
and, thus, its ability to interact, is dominated by gluons.

At small $x$ (high energies) the occupation numbers characterizing
these soft gluon modes are large. This explains the origin of the
main idea of McLerran-Venugopalan (MV) model to describe these
soft gluon modes by tree-level classical Weizsaeker-Williams color
radiation $A^\mu_a$ (the lower index $a$ corresponds to the color
of the gluon mode) of hard partons characterized, in turn, by
static random color charge density $\rho_a$. The physical picture
corresponding to such separation of scales can be described as
follows.

The fast partons, having large longitudinal momenta $p^+$,
propagate along the light-cone emitting and absorbing soft gluons.
In the eikonal approximation this corresponds to having one
nonzero component of the emitting current in the $+$-- direction
$J_a^\mu=\delta^{\mu+}J^+_a$. The hard partons are delocalized in
the longitudinal coordinate $x^-$ at distances $\lambda^- \sim
1/p^+$ and look (almost) pointlike for soft radiation. Of
principal importance is also an hierarchy of temporal scales. For
modes close to the mass-shell one has $2p^+p^- \sim p_\perp^2$, so
that from uncertainty relation the {\it soft} gluons have large
energies (frequencies) $p^-\sim Q^2_\perp/p^+$ and,
correspondingly, short lifetimes $\Delta x^+\sim 1/p^- \sim p^+
\sim x$. At such small lifetimes the dynamics of hard modes is
effectively frozen, so that soft gluons effectively probe static
correlators of hard modes.

The color current describing the hard modes can thus be written as
\begin{equation}\label{jclas}
J_a^\mu(x)=\delta^{\mu+}\rho_a(x^-,{\bf x}_\perp),\,\,\,
\,\,\del^-\rho_a\equiv \frac{\del\rho_a}{\del x^+}=0,\,\,\,\,\,
{\rm supp}\,\rho_a=\{|x^-| \le 1/\Lambda^+ \},\,\,
\end{equation}
In the nonabelian equations of motion describing the tree-level
dynamics of soft glue the current (\ref{jclas}) plays the role of
the source :
\begin{equation} \label{cleq0}
[D_{\nu}, F^{\nu \mu}]\, =\, \delta^{\mu +} \rho_a(x^-,{\bf
x}_\perp)\,.
\end{equation}
The source $\rho_a$ is a stochastic variable with zero mean. The
spatial corelations $\rho_a(\vec x)$ (${\vec x}\equiv (x^-,{\bf
x}_{\perp})$) at the scale $\Lambda^+$ are inherited from
(generally speaking, static) correlators of hard gluons. The
weight of a given charge configuration $\rho_a$ is determined by
some functional $W_\Lambda[\rho]$ which is, by assumption, gauge
invariant. The analysis of the gluon field generated by the source
$\rho_a$ is most transparent in the light-cone gauge $A^+=0$.

The calculation of gluonic correlators  in the MV model proceeds
in two steps:
\begin{itemize}
\item{Solving the classical Yang-Mills equations
 (\ref{cleq0}) in the light-cone gauge $A^+=0$.
 The solution ${\cal A}^i(\vec x)[\rho]$ is some nonlinear functional of $\rho$
 (below we will show that it is always possible to construct a static solution
 of (\ref{cleq0}) having $A^-=0$.)}
\item{Computation of correlators on this classical solution by averaging with respect
 to $\rho$ with the weight $W_\Lambda[\rho]$} :
\end{itemize}
\begin{equation}\label{clascorr}
\langle A^i_a(x^+,\vec x)A^j_b(x^+,\vec y)
\cdots\rangle_\Lambda\,=\, \int {\cal
D}\rho\,\,W_\Lambda[\rho]\,{\cal A}_a^i({\vec x}) {\cal
A}_b^j({\vec y})\cdots\,
\end{equation}
where $\vec x \equiv (x^-,{\bf x}_{\perp})$ and the normalization
of correlators is fixed by
\begin{equation}\label{normF}
\int {\cal D}\rho\,\,W_\Lambda[\rho]\,=\,1.
\end{equation}

It is important to note that the corelators (\ref{clascorr})
depend on the scale  $\Lambda^+$. As will be discussed below, the
effective theory specified by the equations
~(\ref{cleq0})--(\ref{clascorr}) does in fact hold for the modes
having having longitudinal momenta that are not too small as
compared to the reference scale $\Lambda^+$. At very small
longitudinal momenta $b\Lambda^+$ with $b\ll 1$, one has to take
care of the (large) quantum corrections of order of
$\alpha_s\ln(1/b)$. To calculate correlators at the new scale
$b\Lambda^+$, one has to construct a new effective theory through
integration over the quantum degrees of freedom with longitudinal
momenta in the strip $b\Lambda^+ <|p^+|<\Lambda^+$.

\subsection{Classical solution}\label{classol}

Understanding the structure of the classical solution of
(\ref{cleq0}) is a key to the physics of MV model. Before turning
to the analysis of the non-abelian case, it is illuminating to
consider its abelian simplification, i.e. solve the equation
$\partial _{\nu} {\cal F}^{\nu \mu} =  \delta^{\mu +} \rho({\vec
x})$ in the light-cone gauge $A^+=0$. For the sought for static
solution one gets from the $\mu=-$ and $\mu=i$ components of the
equations of motion ${\cal A}^-=0$ (so that ${\cal F}^{-+}={\cal
F}^{i-}=0$) and ${\cal F}^{ij}=0$. The static solution we are
looking for is thus a two-dimensional pure gauge:
\begin{equation} \label{aaimom}
{\cal A}^i(p) \, = \, -{p^i \over p^+} {\rho(p^+,p_{\perp}) \over
p_{\perp}^2}\,.
\end{equation}
To specify the solution completely one has to choose some
prescription for the axial pole $p^+=0$. Let us choose the
prescription $1/ p^+\,\equiv\,1/(p^+\ +i \varepsilon)$. In
coordinate space this leads to the solution of the form
\begin{equation}\label{aiab1}
{\cal A}^i (x^-,x_{\perp})= \int_{-\infty}^{x^-} dy^- \,
\partial^i \alpha (y^-,x_{\perp})\,,
\end{equation}
vanishing at $x^-\to -\infty$. The function $\alpha ({\vec x})$
satisfies $-\nabla_{\perp}^2 \alpha ({\vec x}) = \rho ({\vec x})$.
Different prescriptions for the axial pole correspond to the same
electric field ${\cal F}^{i+}({\vec x})=\partial^i \alpha ({\vec
x})$ and thus to the prescription-invariant physics.

Turning now to the analysis of the non-abelian case, let us note,
that for static charge density $\rho$ equations (\ref{cleq0}) are,
generally speaking, not consistent. Indeed, from the identity
$[D_\mu, [D_{\nu}, F^{\nu \mu}]]=0$ there follows a covariant
conservation of the color current $[D_\mu, J^\mu]=0$, so that the
considered current $J^\mu=\delta^{\mu +}J^+$ should satisfy
$[D^-,J^+]\,\equiv\,\del^- J^+\,-\,ig[A^-,J^+]\,=\,0$ which (at
${\cal A^-} \neq 0$) it does not. The current is static only up to
the isotopic precession $J^+(x^+, \vec x)\,=\,W(x^+,\vec
x)\,\rho(\vec x) \,W^\dagger(x^+,\vec x)$, where $\rho$ is some
initial orientation of the color charge density at some
$x^+=x^+_0$ and $W[A^-]$ is a time-ordered Wilson line
\begin{equation}\label{WLINE1}
     W[A^-](x^+,\vec x)\,\equiv\,{\rm T}\, \exp\left\{\, ig\int_{x^+_0}^{x^+} dz^+ A^-(z^+, \vec x) \right\}.
\end{equation}
Analogously to the above-described abelian case one can, however,
consider the static solution of the form
\begin{equation}\label{Ansatz}
 A^+\,=\,A^-\,=\,0,\qquad A^i\,\equiv\,{\cal A}^i(x^-,x_{\perp})\,.
\end{equation}
The solution (\ref{Ansatz}) is invariant under gauge
transformations independent on $x^-$ and $x^+$, i.e. under
two-dimensional transformations in the transverse plane. Then for
the $\mu=+$ one has $[D_i, F^{i+}] \,=\,\rho({\vec x})$, while for
the  $\mu=i$ one obtains $[D_j, F^{ji}] = 0$, having a
two-dimensional pure gauge solution (${\cal F}^{ji} = 0$):
\begin{equation}\label{tpg}
 {\cal A}^i(x^-,x_{\perp})\,=\,{i \over g}\,
 U(x^-,x_{\perp})\,\partial^i U^{\dagger}(x^-,x_{\perp})\,,
\end{equation}
where $U(x^-,x_{\perp})$ belongs to $SU(N)$ and has an implicit
dependence on $\rho$. The fields ${\cal A}^i$ in~(\ref{tpg}) can
be gauge-rotated to zero by the gauge transformation $U^{\dagger}
({\vec x})$:
\begin{equation}\label{gtr}
 {\cal A}^\mu\longrightarrow {\tilde {\cal A}}^\mu=
 U^{\dagger}{\cal A}^\mu U+ \,{i \over g}\, U^{\dagger} \partial^\mu U.
\end{equation}
leaving the only nonzero component ${\tilde {\cal A}}^+ = {i \over
g}\, U^{\dagger} \left( \partial^+ U \right )$. Note that in this
rotated gauge the gauge potential satisfies the covariant gauge
constraint $\partial_\mu {\tilde A}^\mu =0$. The Yang-Mills
equations take the simple form $- \nabla^2_\perp {\tilde A}^+(\vec
x)\,=\,{\tilde \rho}(\vec x)$, where
\begin{equation}\label{COVRHO}
{\tilde \rho}(\vec x)\,\equiv\,U^{\dagger}(\vec x) \, \rho(\vec x)
\,U(\vec x)
\end{equation}
is a classical color charge in the rotated gauge. It is convenient
to introduce, in the analogy to the abelian case, a new function
$\alpha({\vec x}) \equiv {\tilde {\cal A}}^+({\vec x})$, so that
$\alpha({\vec x})$ satisfies $ - \nabla^2_\perp \alpha({\vec
x})\,=\,{\tilde \rho}(\vec x)$. In computing the gluon correlators
it is useful to use an explicit expression for $U$ in terms of
$\alpha$:
\begin{equation}\label{UTA}
 U^{\dagger}(x^-,x_{\perp})= {\rm P}
 \exp
 \left \{
 ig \int_{x_0^-}^{x^-} dz^-\,\alpha (z^-,x_{\perp})
 \right \},
\end{equation}
where $P$ denotes ordering of the matrices $\alpha(\vec x)$ from
left to right in ascending (descending) order in $x^-$ at
$x^->x_0^-$ ($x^-<x_0^-\,$) correspondingly. Various choices of
$x^-_0$ correspond to solutions related by residual two-dimensonal
gauge transformations. We have thus fully constructed a static
classical solution ${\cal A}^i [\tilde \rho]$ in the light-cone
gauge as an implicit nonlinear functional of the source ${\tilde
\rho}\,$. Explicit construction of the solution is, obviously, not
possible - one would have to explicitly solve for $\alpha$ the
nonlinear equation $U[\rho]\Bigl(-\nabla^2_{\perp} \alpha 
\Bigr) U^{\dagger}[\rho]  \,=\, {\rho}$. For hard modes the source
$\rho$ is localized in the vicinity of $x^-=0$, cf. (\ref{jclas}).

Let us now turn to the all-important issue of fixing the residual
gauge invariance. To do it at the tree level ( i.e. at the
classical solution (\ref{tpg})), we shall again use the retarded
boundary conditions in   $x^-\,$: ${\cal A}^i(\tilde x)\to 0$ for
$x^- \to -\infty\,$, which is equivalent to choosing $x^-_0\to
-\infty$ in (\ref{UTA}). The choice of this boundary condition
fixes, in fact, the axial pole prescription for the gluon
propagator used in computing the quantum corrections. Note also,
that the chosen retarded prescription corresponds to the source
having its support only at positive $x^-$ in the interval $0\simle
x^-\simle 1/\Lambda^+$.

In MV approach $\Lambda^+$ is a big longitudinal scale, and both
the size of the probe and the longitudinal scale characterizing
the soft gluon fields satisfy $p^+\ll \Lambda^+$ so that these
fields  resolve only the rough longitudinal structure of the
localized source. This allows to simplify the formula for the
classical solution at parametrically large distances from the
source by using the following approximation for the rotation
matrices
 \begin{equation}\label{UTAF}
U^{\dagger}(x^-,x_{\perp})\,\equiv\,
 {\rm P} \exp
 \left \{
 ig \int_{-\infty}^{x^-} dz^-\,\alpha (z^-,x_{\perp})
 \right \}\approx \,\theta(x^-)\,\Omega^\dagger(x_{\perp}) + \theta(-x^-),
\end{equation}
where
\begin{equation}\label{V}
 \Omega^\dagger(x_{\perp})\,\equiv\,{\rm P} \exp
 \left \{
 ig \int_{-\infty}^{\infty} dz^-\,\alpha (z^-,x_{\perp})
 \right \}.
\end{equation}
so that(cf.~(\ref{tpg})):
\begin{equation}\label{APM}
 {\cal A}^i(x^-,x_\perp)\,\approx\,\theta(x^-)\, \frac{i}{g}\,\Omega(\del^i
 \Omega^\dagger) \,\equiv\,\theta(x^-){\cal A}^i_\infty(x_\perp),
\end{equation}
and the chromoelectric field strength is effectively a
delta-function\footnote{Note that $\delta$-- и $\theta$--
functions in these formulae are understood as being regularized at
distances $\Delta x^-\sim 1/\Lambda^+$.}:
\begin{equation}\label{FDELTA}
 {\cal F}^{i+}(\vec x) \,\equiv\, -\partial^+{\cal
A}^i\,\approx\,-\delta(x^-)\, {\cal A}^{i}_\infty(x_\perp).
\end{equation}

Let us remind the standard definition of the gluon distribution
function in the light-cone gauge $A^+=0$:
 \begin{eqnarray}\label{GDF0}
  G(x,Q^2)&\equiv& \int {d^2k_\perp \over (2 \pi)^2}\,\Theta(Q^2-
  k_\perp^2)\int{dk^+ \over 2 \pi} \,2k^+\,\delta\biggl(x-{k^+\over P^+}\biggr)\nonumber\\
  &{}&\qquad\qquad \Bigl\langle
  A^i_a(x^+,k^+,{\bf k}_\perp) A^i_a(x^+,-k^+,-{\bf k}_\perp)\Bigr\rangle,\,\,\,
 \end{eqnarray}
where the averaging is over the wave function of hadron (nucleus).
The equation  ~(\ref{GDF0}) can be interpreted as follows. In the
theory quantized on the light-cone  ($\vec k \equiv (k^+,{\bf
k}_\perp)$) the expression
 \begin{equation}\label{lcc}
  \frac{2k^+}{(2 \pi)^3}\, A^i_c(x^+,\vec k) A^i_c(x^+,-\vec k)\,=\,
  \sum_\lambda\sum_c a^\dagger_{\lambda c}(\vec k)\,a_{\lambda c}(\vec k)\,=\,
  \frac{dN}{d^3 k}\,
 \end{equation}
corresponds to gluon density in Fock space, i.e. the number of
gluons with given momentum in unit volume \footnote{In Eq.
(\ref{lcc}) $a^\dagger_{\lambda c}(\vec k)$ и $a_{\lambda c}(\vec
k)$ are creation and annihilation operators for gluons with
momentum $\vec k$, color $c$ and transverse polarization
$\lambda$.}. The definition  ~(\ref{GDF0}) corresponds, therefore,
to counting the number of gluons with longitudinal momentum
$k^+=xP^+$ and transverse one  $k_\perp \leq Q$ in hadron
(nuclear) wavefunction.

At tree level one gets
 \begin{equation}\label{GCL}
  x G_{cl}(x,Q^2)\,=\,\frac{1}{\pi} \int {d^2k_\perp \over (2
  \pi)^2}\,\Theta(Q^2- k_\perp^2)\,\Bigl\langle\, |{\cal
  F}^{i+}_a(\vec k)|^2\Bigr\rangle_\Lambda\,,
 \end{equation}
where averaging over $\rho$ is performed at the scale
$\Lambda^+\sim xP^+$ (see ~(\ref{clascorr})--(\ref{normF})). Using
~(\ref{FDELTA}) for ${\cal F}^{i+}$, we obtain
 \begin{eqnarray}\label{GCL1}
  x G_{cl}(x,Q^2)&=&\frac{1}{\pi} \int {d^2k_\perp \over (2
  \pi)^2}\,\Theta(Q^2- k_\perp^2)\,\Bigl\langle\, |{\cal
  F}^{i+}_a(k_\perp)|^2\Bigr\rangle_\Lambda\nn &=&{R^2} \int^{Q^2}
  {d^2k_\perp \over (2 \pi)^2} \int d^2x_\perp\,{\rm
  e}^{-ik_\perp\cdot x_\perp} \Bigl\langle  {\cal
  A}^{ia}_\infty(0)\,
  {\cal A}^{ia}_\infty(x_\perp)\Bigr\rangle_\Lambda,
 \end{eqnarray}
where $R$ is a radius of the hadron (nucleus) and we assume, for
simplicity, homogeneity in the impact parameter plane. For
momentum density of gluons in the transverse plane one gets
 \begin{equation}\label{WIGG}
  N(k_\perp)\,\equiv\,\frac{d^2(x
  G_{cl})}{d^2k_\perp\,d^2b_\perp}\,\equiv\, \int d^2x_\perp\,{\rm
  e}^{-ik_\perp\cdot x_\perp} \Bigl\langle  {\cal
  A}^{ia}_\infty(0)\, {\cal A}^{ia}_\infty(x_\perp)\Bigr\rangle,
 \end{equation}

Note that in the considered approximation the dependence of the
gluon density on $x$ is only due to the $x$ - dependent weight
functional $W_\Lambda[\rho]$\, ($\Lambda^+\sim xP^+$), so that in
the MV model the whole dependence on $x$ is encoded in the weight
functional and is ultimately determined by its (quantum)
evolutionary dependence on $\Lambda^+$.

In the linear approximation in $\rho$ we have ${\cal F}^{+j}_a
\simeq i(k^j/k^2_\perp)\rho_a\,$ and, thus,
 \begin{equation}\label{GCLLIN}
  xG_{cl}(x,Q^2)\,\simeq\,\frac{1}{\pi} \int {d^2k_\perp \over (2
  \pi)^2}\, \frac{\Theta(Q^2-k_\perp^2)}{k^2_\perp}\,\Bigl \langle
  |\,\rho_a(\vec k)|^2\Bigr\rangle_\Lambda.\,
 \end{equation}

\subsection{Gluon distribution in MV model of the nucleus: low
density limit}

The simple model of color source generating the gluonic component
of the nuclear wavefunction proposed by McLerran and Venugopalan
\cite{MV94} corresponds to considering the $A \times N_c$
constituent quarks in the nucleus as an ensemble of {\it
independent} color sources. The main approximation made in this
model is, clearly, a neglect of correlations between the colors of
constituent quarks belonging to the same nucleon due to
confinement, which for small enough probes and large enough
nucleus should be a good approximation. The total color charge in
the tube having transverse cross-section $\Delta S_\perp$ is
described by its moments
 \begin{equation}\label{QQQ}
  \langle Q^a \rangle = 0, \,\,\,\,\,\,\,
  \langle Q^a Q^a \rangle = \Delta S_\perp
  {g^2 C_f N_c A \over \pi R_A^2} \equiv
  \Delta S_\perp {g^2 (N^2_c-1) A \over 2 \pi R^2_A}
 \end{equation}
so that for the color charge density one has
 \begin{equation}\label{cd0}
  {\langle Q^a Q^a \rangle \over \Delta S_\perp (N^2_c-1)} =
  {\alpha_s \over 2} {A \over \pi R^2_A} \equiv \mu^2_A
 \end{equation}
In terms of the color charge density $\rho^a(x^-,x_\perp)$
 \begin{equation}
  Q^a \, = \, \int_{\Delta S_\perp} d^2 x_\perp \int dx^-
  \rho^a(x^-,x_\perp) \equiv \int_{\Delta S_\perp} d^2 x_\perp
  \rho^a(x_\perp)
 \end{equation}
the moments of the color charge distribution (\ref{QQQ})
correspond, assuming homogeneity in the impact parameter plane, to
the following correlators
 \begin{eqnarray}\label{rhorho}
  \langle \rho^a(x^-,x_\perp) \rho^b(y^-,y_\perp) \rangle & = &
  \delta^{ab} \delta (x_\perp - y_\perp) \delta (x^- - y^-)
  \lambda_A (x^-) \nonumber \\
  \langle \rho^a(x_\perp) \rho^b(y_\perp) \rangle & = & \delta^{ab}
  \delta (x_\perp - y_\perp) \mu_A
 \end{eqnarray}
or, in momentum space,
 \begin{equation}\label{rhorhom}
  \langle \rho^a(k_\perp) \rho^a (-k_\perp) \rangle \, = \, \pi R^2_A \mu_A^2
 \end{equation}
where $\mu^2_A \equiv \int dx^- \lambda_A (x^-) = g^2 A/2 \pi
R^2_A$ (see (\ref{cd0})). The weight functional generating the
correlators (\ref{rhorho}) is, evidently, a gaussian
 \begin{equation}\label{mvwf}
  W_A[\rho] \, \sim \, \exp \left [ - \frac{1}{2} \int d^3 x
  {\rho_a ({\vec x}) \rho_a (\vec x) \over \lambda_A (x^-)} \right ].
 \end{equation}
Using (\ref{GCLLIN}) and (\ref{rhorho}) one obtains final
expressions for the gluon momentum density in the transverse plane
${\cal N}_A (k_\perp)$ and gluon structure function
$xG_{cl}(x,Q^2)$
 \begin{eqnarray}\label{linan}
  {\cal N}_A (k_\perp) & \simeq & {N^2_c-1 \over 4 \pi^3} {\mu^2_A
  \over k_\perp^2} \equiv {N^2_c-1 \over 4 \pi^3}\, \varphi_A (k_\perp),\nonumber \\
  xG_{cl}(x,Q^2) & \simeq & {N^2_c-1 \over 4 \pi} R^2_A \mu^2_A
  \int_{\Lambda^2_{QCD}}^{Q^2} {dk^2_\perp \over k^2_\perp} =
  A N_c {\alpha_s C_F \over \pi} \ln {Q^2 \over \Lambda^2_{QCD}}\,,
 \end{eqnarray}
where in the first line of Eqn.~(\ref{linan}) we introduced a
so-called unintegrated structure function $\varphi$, and in the
second one immediately recognizes the standard lowest order
perturbative spectrum of gluons radiated by $AN_c$ quarks in the
nucleus.

\subsection{Gluon distribution in MV model of the nucleus:
saturation}

In the previous subsection we have calculated the gluon density in
the transverse plane ${\cal N}_A (k_\perp)$ in the low density
regime. Technically "low density" meant neglecting the non-abelian
effects in computing the correlation of chromoelectric fields (cf.
Eqns. (\ref{GCL},\ref{GCL1},\ref{WIGG}) ). The fully non-abelian
calculation was performed in \cite{JKMW97,KM98}, see also a
detailed and transparent derivation in \cite{IV03}. The answer for
the gluon density in the transverse plane ${\cal N}$ reads
 \begin{equation}\label{satden}
  {\cal N}_A (k_\perp) \, = \, {N^2_c-1 \over 4 \pi^4 \alpha_s N_c}\,
  \int {d^2 x_\perp \over x^2_\perp} {\rm e}^{-ik_\perp \cdot x_\perp}
  \left [ 1-\exp \left ( -{1 \over 4} x_\perp^2 Q^2_A\,
  \ln {1 \over x^2_\perp \Lambda^2_{QCD}} \right ) \right ]
 \end{equation}
where
 \begin{equation}\label{satmom1}
  Q^2_A \equiv \alpha_s N_c \mu^2_A = \alpha_s N_c \int dx^- \lambda_A (x^-)
 \end{equation}
From (\ref{satden}) it is clear, that the (transverse) momentum
scale $Q^2_s$, at which nonlinear effects become important is
determined by the following nonlinear equation:
 \begin{equation}\label{satmom2}
  Q^2_s \, \simeq \, Q^2_A \, \ln {Q^2_s \over \Lambda^2_{QCD}}\,,
 \end{equation}
where the characteristic transverse distance was taken to be
$1/r^2_\perp=Q^2_s$. The properties of the nonlinear gluon
distribution (\ref{satden}) are best illustrated by considering
its low-density (high transverse momentum) and high density (low
transverse momentum) asymptotics. Expressed the in terms of the
unintegrated structure function $\varphi$ introduced in Eqns.
(\ref{linan})), (\ref{satden}) interpolates between the following
asymptotics:
 \begin{eqnarray}\label{satdenas}
  \varphi_A (k_\perp \gg Q_s) = {\mu_A^2 \over k^2_\perp} \,\,\, \to \,\,\,
  \varphi_A (k_\perp \ll Q_s) = {1 \over \alpha_s N_c }\, \ln {Q^2_s \over k^2_\perp}
 \end{eqnarray}
Equation (\ref{satdenas}) demonstrates the key property of
(\ref{satden}): the gluon density {\it saturates} at small
momenta, replacing a powerlike infrared divergence of perturbative
asymptotics by a mild logarithmic divergence. The transverse
momentum scale $Q_s$ that controls this transition is,
appropriately, called saturation momentum.

In the following we shall need an expression for the gluon nuclear
number density in the model, where the nucleus is composed of
colorless combinations of constituent quarks ("nucleons"). More
precisely, we are interested in the transverse phase space density
at midrapidity
\begin{equation}\label{GCLNLIN}
 {d N_g \over dy db_\perp dk_\perp} \equiv
 {d (x G_A) \over db_\perp dk_\perp} \, = \,
 {N^2_c-1 \over 4 \pi^4 \alpha N_c} \int {dx_\perp \over x_\perp^2}\
 {\rm e}^{-ik_\perp x_\perp} \left (
 1-{\rm e}^{x^2_\perp Q^2_s/4} \right )
\end{equation}
where the saturation momentum (considered at given impact
parameter $b_\perp$) is related to the nucleon structure function
$x G_{nucleon}(x,Q^2_s)$ through
\begin{equation}\label{satmom}
Q^2_s(b,x) \, = \, {4 \pi^2 \alpha N_c \over N^2_c-1}\, 2\,
\sqrt{R^2-b^2}\, \rho \, x G_{nucleon} (x,Q^2_s)
\end{equation}

\section{Quantum Corrections in the High Energy Limit}

In the previous section we have discussed the McLerran-Venugopalan
approach to high energy QCD description of dense partonic systems
at tree (classical) level. In the present section we describe a
systematic approach to computing the quantum corrections to the
tree-level description.

\subsection{Renormalization group}

The physics of quantum corrections to the tree-level MV picture is
that of the quantum modes with longitudinal momenta
$|p^+|<\Lambda^+$ considered in addition to the classical modes
${\cal A}^i$ generated by the source $\rho$. The restriction
$|p^+|<\Lambda^+$ comes from the fact that, by assumption, the
modes having $|p^+|>\Lambda^+$ were already integrated out in the
process of construction of the effective theory at the scale
$\Lambda^+$.

The basic object of the theory under construction is a generating
functional of the correlators of gluon fields having longitudinal
momenta in the interval $|p^+|<\Lambda^+$ in the light-cone gauge
$A^+=0$:
\begin{equation}\label{PART}
 {\cal Z}[j]\,=\,\int {\cal D}\rho\,\,W_\Lambda[\rho]
 \,\,Z_{\Lambda}^{-1}[\rho]\int^\Lambda {\cal D}A_a^\mu\,
 \delta(A^+_a)\,\,{\rm e}^{\,iS[A,\,\rho]-i\int j\cdot A}\, \equiv
 \int {\cal D}\rho\,\,W_\Lambda[\rho]\,Z_{\Lambda}[\rho,j].
\end{equation}
Equation (\ref{PART}) includes two functional integrations: over
$A^\mu\,$
\begin{equation}\label{PARTQUANT}
 Z_{\Lambda}[\rho,j]\,\equiv\, Z_{\Lambda}^{-1}[\rho]\int^\Lambda
 {\cal D}A_a^\mu\, \delta(A^+_a)\,\,{\rm e}^{\,iS[A,\,\rho]-i\int
 j\cdot A}\,
\end{equation}
where $Z_{\Lambda}[\rho]\equiv Z_{\Lambda}[\rho,j=0]$ describes
quantum fluctuations at fixed $\rho$, and the "classical"
averaging over $\rho$ with the weight $W_\Lambda[\rho]$. The
subscript "$\Lambda$" denotes integration over the modes having
$|p^+| < \Lambda^+$ \footnote{Note that in the light-cone gauge
the "longitudinal" separation of degrees of freedom is defined
uniquely: the residual gauge transformations do not depend on
$x^-$ and, therefore, can not change the longitudinal momentum
$p^+$.}.

The quantum dynamics of the problem is specified by the effective
action $S[A,\rho]$ such that, first, in the regime $\delta
S/\delta A^\mu=0$ the tree-level equations of motion are
reproduced and, second, a correct quantum evolution of the
correlators of the theory is ensured. Such effective action was
proposed in \cite{JKLW97,JKLW99a,JKW99} (see also \cite{JJV00}):
\begin{equation}\label{ACTIONR}
 S[A,\rho]\,=\,- \int d^4x \,{1 \over 4} F_{\mu\nu}^a F^{\mu\nu}_a
 \,+\,{i \over {gN_c}} \int d^3 \vec x\, {\rm Tr}\,\Bigl\{
 \rho(\vec x) \,W_{\infty,-\infty}(\vec
 x)\Bigr\}\,\equiv\,S_{YM}\,+\,S_W.,\,\,
\end{equation}
where
\begin{equation}\label{WLINER}
 W_{\infty,-\infty}[A^-](\vec x)\, =\,{\rm T}\, \exp\left[\, ig\int
 dx^+ A^-(x) \right].
\end{equation}
Effective action (\ref{ACTIONR}) includes the standard Yang-Mills
piece $S_{YM}$, as well as a gauge invariant generalization of the
abelian eikonal vertex  $\int d^4x \,\rho_a A^-_a$
\footnote{Strictly speaking, because of the nonlocal dependence of
the eikonal interaction of source $\rho$ with $A^-$ on time,
effective action should be considered on the contour in the
complex plane. It can be shown, however, that in the leading
logarithmic approximation one could restrict oneself to
considering the dynamics on the real time axis \cite{ILM00}.}.

At tree level $A^\mu_a\approx {\cal A}^\mu_a =\delta^{\mu i} {\cal
A}^i_a$, where ${\cal A}^i_a$ is a solution of the classical
equations of motion (EOM) with source $\rho_a$ described above.
The full gluon field in Eqn. (\ref{PART}),(\ref{PARTQUANT})
includes both classical and quantum components:
\begin{equation}\label{fluct0}
 A^\mu_a(x)\,=\,{\cal A}^\mu_a(x) + \delta
 A^\mu_a(x).
\end{equation}
The mean field  $\langle A^\mu_a (x)\rangle$ includes ${\cal
A}^\mu$, as well as the contribution induced by quantum
fluctuations $\langle \delta A^\mu\rangle$ corresponding to
polarization of gluon fluctuations by the external charge
\begin{equation}
 \langle A^\mu_a (x)\rangle\,=\, {\cal A}^\mu_a(x) + \langle \delta
 A^\mu_a(x)\rangle\,\equiv\, {\cal A}^\mu_a(x) + \delta {\cal
 A}^\mu_a(x)
\end{equation}
and satisfies (the brackets denote quantum averaging at fixed
$\rho$):
 \begin{equation}\label{MFE}
 \left\langle \frac{\delta S}{\delta A^\mu_a(x)}
 \right\rangle\,=\,0.
\end{equation}

Using the generating functional (\ref{PART}), one could compute
arbitrary gluon correlators. For example, the two-point correlator
reads (double brackets indicate averaging with respect to both
quantum fluctuations and external source)
\begin{equation}\label{2point}
 << {\rm T}\,A^\mu(x)A^\nu(y) >> \,\,=\,\int {\cal
 D}\rho\,\,W_\Lambda[\rho] \left\{\frac{\int^\Lambda {\cal D}A
 \,\,A^\mu(x)A^\nu(y)\,\,{\rm e}^{\,iS[A,\,\rho]}} {\int^\Lambda
 {\cal D}A\,\,{\rm e}^{\,iS[A,\,\rho]}} \right\}.
\end{equation}

The sought for effective theory can be constructed by
layer-by-layer integration of quantum fluctuations over $p^+$ (or
$p^-$). The dominating contributions at small $x$ are those
proportional to large rapidity intervals $\Delta \tau=\ln(1/x) \gg
1$. We will see, that integration over $p^+$ in the strip $k^+\ll
p^+\ll\Lambda^+$ generates corrections of order
$\alpha_s\ln(\Lambda^+/k^+)$ to amplitudes with external momenta
$k^+ < \Lambda^+$ which are essential if $\Lambda^+\gg k^+$. In
the picture including quantum fluctuations MV model describes
correlations of gluon fields at tree level, if all degrees of
freedom with longitudinal momenta exceeding $k^+$ are integrated
out, with corresponding induced contributions included into
parameters of the action. Potentially large logarithmic
contributions are resummed into quantum evolution of the weight
functional $W_\Lambda[\rho]$, where $\Lambda^+\sim k^+$. The
resulting classical theory is valid at the scale $k^+$ (the
contributions of higher order in $\alpha_s$ are neglected).

\subsection{Linear evolution: BFKL limit}

The law of the evolution of gluon density $\cal N$ with energy is
given, in the leading logarithmic approximation, by the BFKL
equation \cite{BFKL}. Its coordinate representation has the
following form (see, e.g., \cite{IV03}):
$$
  {\partial {\cal N} (x_\perp,y_\perp) \over \partial \tau} \, = \,
  - {\alpha_s \over \pi} \int d^2z_{\perp} {(x_\perp-y_\perp)^2
  \over (x_\perp-z_\perp)^2(y_\perp-z_\perp)^2 }
 $$
 \begin{equation}\label{cBFKL}
    \times  \left [ {\cal N} (x_\perp, z_\perp) + {\cal N} (z_\perp, y_\perp) -
    {\cal N} (x_\perp, y_\perp) \right ] _{\tau}
 \end{equation}
The key property of the solutions of Eq.~(\ref{cBFKL}) is their
exponential growth with $\tau$,
 \begin{equation}
  {\cal N} (\tau \to \infty) \sim {\rm e}^{c \tau},
 \end{equation}
which means, that physical cross-sections calculated in the linear
approximation in gluon density have a powerlike divergence in
energy in the high energy limit, violating unitarity. Thus, the
linear density approximation is conceptually unsatisfactory and
has to be ameliorated. A natural possible way of achieving such
improvement is to construct a consistent nonlinear generalization
of the linear formalism. This is the logical line that we shall
follow below.

\subsection{Nonlinear evolution equation}

To describe the quantum evolution of the weight functional
$W_\Lambda[\rho]$ with $\Lambda^+$, it is convenient to introduce
two theories, "Theory I" and "Theory II", which differ by their
longitudinal scales --  $\Lambda^+$ and $b\Lambda^+$
correspondingly, where $b\ll 1$, but $\alpha_s\ln(1/b) < 1$. In
Theory II the modes in the strip
 \begin{equation}\label{strip}
  \,\, b\Lambda^+ \,\,<\,\, |p^+|\,\, <\,\,\Lambda^+\,
 \end{equation}
separating Theories I and II are integrated out, and induced
contributions resulting from this integration are taken into
account in the weight functional $W_{b\Lambda}$ by suitable
redefinition of its coefficients.

\subsubsection{Nonlinear evolution equation: derivation}

To explicitly calculate $\Delta W \equiv W_{b\Lambda} - W_\Lambda$
one should compare expressions for gluon correlators calculated at
a scale $k^+\simle b\Lambda^+$ in both theories. In Theory II, in
he leading order in $\alpha_s$, induced effects are present at
tree level. In Theory I one has logarithmically amplified
contributions from quantum fluctuations in the strip
(\ref{strip}). In computing quantum corrections we shall keep the
terms of the leading order in  $\alpha_s\ln(1/b)$ (Leading
Logarithmic Approximation -- LLA ), but of all orders in
background fields and sources. This is necessary because of the
key role of strong fields ${\cal A}^i \sim 1/g$ and sources
$\rho\sim 1/g$ in our problem. The resulting equation
\cite{JKLW97,JKLW99a} is a nonlinear functional equation on
$W_\tau[\rho]$ (here $\tau\equiv \ln(1/b)$), describing the
evolution of $W_\tau[\rho]$ with $\tau$.

Let us schematically consider calculations in Theory I at an
example of the two-point equal time correlator $ \langle
A^i_a(x^+,\vec k) A^i_a(x^+,-\vec k)\rangle $ or, more precisely,
its coordinate counterpart ${\cal G}(\vec x, \vec y) \,\equiv\,
\langle A^i_a(x^+,\vec x) A^i_a(x^+,\vec y)\rangle\ $, which is,
in fact, independent of $x^+$ because of the static nature of the
source. In what folows it is convenient to introduce the following
decomposition of the full gluon field:
 \begin{equation}\label{fluct1}
  A^\mu_a(x)\,=\,{\cal A}^\mu_a(x) + \delta A^\mu_a(x) + a^\mu_a(x).
 \end{equation}
where ${\cal A}^\mu_a(x)$ is a classical solution, $\delta
A^\mu_a(x)$ are semihard quantum fluctuations  with longitudinal
momenta in the strip (\ref{strip}) and $a^\mu_a(x)$ are soft
fields. Quantum effects important for our calculation arise from
interaction of the soft modes $a^{\mu}$ with semihard ones $\delta
A^\mu$ in the presence of external field ${\cal A}^i$ and source
$\rho$. Detailed analysis shows \cite{ILM00,FILM02} that in the
leading logarithmic approximation  $\delta {\cal A}^i \sim
\alpha_s\,{\rm log}(1/b){\cal A}^i$ and $\langle a^i a^i \rangle
\sim \alpha_s\,{\rm log}(1/b){\cal A}^i {\cal A}^i$ so that
 \begin{equation}\label{2PL}
  {\cal G}(\vec x, \vec y) = {\cal
  A}^i(\vec x){\cal A}^i(\vec y) + {\cal A}^i(\vec x)\delta {\cal
  A}^i(\vec y) + \delta{\cal A}^i(\vec x){\cal A}^i(\vec y)+ \langle
  a^i(x^+,\vec x) a^i(x^+,\vec y)\rangle.
 \end{equation}
The correlator (\ref{2PL}) contains three principal contributions:
the tree level one, induced mean field and induced density
corresponding to gluon polarization in the presence of the
external source. The smallness of contributions nonlinear in $a^i$
is due to the smallness of induced fields as compared to
background ones. From ${\cal A}^i$ and $\rho$ being static it
follows that the induced mean field $\delta{\cal A}^i$ is static
as well, and the two-point functions, such as  $\langle a^i_x
a^j_y \rangle$ depend only on $x^+-y^+$. Let us also note that
$\langle \delta A^\mu\rangle=0$.

As mentioned above, in calculating $\delta {\cal A}^i$ and
$\langle\delta A^i\delta A^i\rangle$ we shall be looking at
quantum contributions coming from semihard gluons. Let us note
that leading logarithmic approximation imposes stringent
restrictions on the kinematical definition of the semihard modes
-- these are the near-mass shell ones with longitudinal momenta
$b\Lambda^+\ll |p^+|\ll \Lambda^+$ and frequencies $\Lambda^-\ll
|p^-|\ll \Lambda^-/b$, where
 \begin{equation}\label{strips}
  \Lambda^-\,\equiv\,\frac{Q_\perp^2}{2\Lambda^+}\,,
 \end{equation}
where $Q_\perp$ is some characteristic transverse momentum.

The aim of the calculation is to express $\delta {\cal A}^i$ and
$\langle a^i a^i\rangle$ through correlators of semihard gluon
modes to one-loop accuracy in LLA in $\alpha_s\ln(1/b)$.
Corresponding interaction vertices can be read off the expansion
of the effective action $S[{\cal A}+ \delta A + a]$ to quadratic
order in $\delta A$. In the approximation used it is sufficient to
consider the contributions of the form $a^\mu_a \delta J_\mu^a\,$,
where
 \begin{eqnarray}\label{deltaJ}
  \, \delta J_\mu^a (x) &\equiv&-\,\frac{\delta S}{\delta
  A^\mu_a(x)} \bigg|_{{\cal A}+\delta A}\\
  &\approx&-\,\frac{\delta^2 S}{\delta A^\mu_a(x)\delta
  A^\nu_b(y)}\bigg|_{ {\cal A}}\,\delta A^\nu_b(y)\,-\,\frac{1}{2}\,
  \frac{\delta^3 S}{\delta A^\mu_a(x)\delta A^\nu_b(y) \delta
  A^\lambda_c(z)}\bigg|_{ {\cal A}}\,\delta A ^\nu_b(y) \delta A
  ^\lambda_c(z)\,.\nonumber
 \end{eqnarray}

Let first consider the last contribution to Eq. (\ref{2PL}).
Generically
 \begin{equation}\label{JJSIG}
  \langle a^\mu_a(x) a^\nu_b(y)\rangle\,=\,\int d^4z \int d^4u\,
  G^{\mu\alpha}_{R\,ac}(x,z)\,\langle\delta J_\alpha^{c} (z) \delta
  J_\beta^{d} (u)\rangle\,G^{\beta\nu}_{A\,db}(u,y),
 \end{equation}
where $G_R$ ($G_A$) are retarded (advanced) propagators of the
soft field $a$ in the presence of background field ${\cal A}$ and
source $\rho$ calculated by inverting the operator
 \begin{equation}\label{invG}
  G_{\mu\nu}^{-1}(x,y)[{\cal A},\,\rho]\,\equiv\,
  \frac{\delta^2 S [A,\,\rho]} {\delta
  A^\mu(x)\delta A^\nu(y)}\bigg|_{{\cal A}}\,
 \end{equation}
on the subspace of soft modes. In LLA the only contribution to
$\langle\delta J^\mu \delta J^\nu \rangle$ is coming from the
correlator of charge density fluctuations
 \begin{equation}\label{CHIDEF}\,
  \hat\chi_{ab}(x,y)\,\equiv\,\langle\delta \rho_a(x)\, \delta
  \rho_b(y)\rangle\,.
 \end{equation}

Let us now consider the induced mean field contribution $\delta
{\cal A}^\mu$ corresponding to solving (\ref{MFE}) to the leading
order in the coupling constant:
 \begin{equation}\label{MFE1}
  0\,=\,\left\langle \frac{\delta S}{\delta A^\mu_a(x)}\bigg|_{
  {\cal A}+\delta A+ a} \right\rangle\Bigg|_{\langle \delta A
  \rangle =0}\,\approx\, \frac{\delta^2 S}{\delta A^\mu_a(x)\delta
  A^\nu_b(y)}\bigg|_{ {\cal A}} a_b^\nu(y)\,-\,{\cal
  J}_\mu^{\,a}(\vec x),
 \end{equation}
where ${\cal J}_\mu^{\,a}(\vec x)$ denotes induced current
 \begin{equation}\label{JTOT}
  {\cal J}_\mu^{\,a}(\vec x)\equiv-\,\frac{1}{2}\,
  \frac{\delta^3 S}{\delta A^\mu_a(x)\delta A^\nu_b(y) \delta
  A^\lambda_c(z)}\bigg|_{ {\cal A}} \biggl( \Bigl\langle \delta A
  ^\nu_b(y) \delta A^\lambda_c(z) \Bigr\rangle + \Bigl\langle
  a^\nu_b(y) a^\lambda_c(z) \Bigr\rangle\biggr) \,,
 \end{equation}
containing contributions of the two types:

({\it a\/}) The contribution proportional to quantum averaging
$\langle \delta J^\mu_a\rangle$ of the current (\ref{deltaJ}):
 \begin{equation}\label{JIND}
  \, \hat\sigma_a(\vec x)\,\equiv\,\langle\delta \rho_a(x)\rangle\,
  ={\rm O}\bigl(\alpha_s\ln(1/b)\rho\bigr).
 \end{equation}

({\it b\/}) The contribution including the induced correlator
$\langle a^\nu a^\lambda \rangle$ cf. ~(\ref{JJSIG}). To LLA
accuracy the second term in ~(\ref{JTOT}) includes only correlator
of transverse modes $\langle a^i a^j\rangle$ proportional to
$\hat\chi$:
 \begin{equation}\label{delcalJ}
  \delta{\cal J}_\mu\,\equiv\, -\,\frac{1}{2}\, \frac{\delta^3
  S}{\delta A^\mu \delta A^i \delta A^j}\bigg|_{ {\cal
  A}}\Bigl(G^{i-}\hat\chi \,G^{-j}\Bigr),
 \end{equation}
which is of order $g\hat\chi \sim g\alpha_s\ln(1/b)\rho\rho\,$.
For $g\rho \sim 1$ this contribution is of the same order as
$\hat\sigma$.


Thus ${\cal J}^\mu =\delta^{\mu+}\hat\sigma +\delta{\cal
J}^\mu\,$. Because of ${\cal J}^\mu$ being static the solution of
~(\ref{MFE1}) takes the form
 \begin{eqnarray}\label{deltaCA}
  \delta{\cal A}^+_a&=& \delta{\cal A}^-_a\,=\,0\,,\nn \delta{\cal
  A}^i_a(\vec x)&=& \int d^3\vec y\,\,G_{ab}^{\,i\nu}(\vec x,\vec
  y,p^-=0)\, {\cal J}_\nu^{\,b}(\vec y),
 \end{eqnarray}
so that the formula for the gluon correlator (\ref{2PL}) reads
 \begin{equation}\label{2PL1}
  {\cal G}(\vec x, \vec y)= {\cal A}^i_{\vec x}{\cal A}^i_{\vec y}+
  (G^{\,i\nu}{\cal J}_\nu^{})_{\vec x}\,{\cal A}^i_{\vec y}+ {\cal
  A}^i_{\vec x}\,(G^{\,i\nu}{\cal J}_\nu^{})_{\vec y}+
  (G^{\,i-}\hat\chi \,G^{\,-i})_{\vec x\,\vec y}\,,
 \end{equation}
The kernels $\hat\sigma$ (hidden in ${\cal J}^\mu$) and $\hat\chi$
contain the sought for logarithmic enhancement. To the LLA
accuracy one can make the following replacements
 \begin{equation}\label{sighatperp}
  \hat\sigma_a(\vec x)\, \to
  \, \delta(x^-)\, \alpha_s\ln{1\over b}\,\sigma_a ({x}_\perp)
  \equiv \delta(x^-)  \int dx^- \,\hat\sigma_a(x^-, x_\perp)
 \end{equation}
and
 \begin{equation}\label{chihatperp}
  \hat\chi_{ab}(\vec x,\vec y) \to
  \delta(x^-)\,\alpha_s\ln{1\over b}\, \chi_{ab}(x_\perp, y_\perp)\,\delta(y^-),
  \equiv \delta(x^-)\,\int dx^-\int dy^-\,\hat\chi_{ab}(\vec x,\vec y)
 \end{equation}
so that for the correlator (\ref{2PL}) we finally obtain
 \begin{equation}\label{2PL2}
  {\cal G}(\vec x, \vec y)\,\approx\, {\cal A}^i_{\vec x}{\cal
  A}^i_{\vec y}\,+\,\alpha_s\ln(1/b) \Bigl\{ (G^{\,i\nu}{\cal
  J}_\nu)_{\vec x}\,{\cal A}^i_{\vec y}+ {\cal A}^i_{\vec
  x}\,(G^{\,i\nu}{\cal J}_\nu)_{\vec y}+ (G^{\,i-}\chi
  \,G^{\,-i})_{\vec x\,\vec y} \Bigr\},
 \end{equation}
where the logarithmic enhancement $\ln(1/b)$ is shown explicitly.
After averaging over $\rho$ with weight functional
$W_{\Lambda}[\rho]$, ~(\ref{2PL2}) describes the gluon density at
the scale $b\Lambda^+$ calculated to the LLA accuracy in Theory I.

The kernels $\sigma$ and $\chi$ can be computed analytically
\cite{JKW99,ILM00,FILM02,ILM02}. The result seems to depend on the
technical assumptions used in the calculations described in
\cite{JKW99} and \cite{ILM00,FILM02,ILM02}
 (see also \cite{L00}). The fact that the form of nonlinear terms in
QCD evolution equations depends, in particular,  on residual gauge
fixing is in fact not new, see e.g. \cite{MQ86}.

Let us now calculate the same gluon correlator (\ref{2PL}) in
Theory II. By construction
 \begin{equation}\label{evolG}
  \langle{\cal A}^i_{\vec x}{\cal A}^i_{\vec y}
  \rangle_{b\Lambda}=\langle{\cal A}^i_{\vec x}{\cal A}^i_{\vec y}
  \rangle_{\Lambda}+ \alpha_s\ln(1/b)\Bigl\langle (G^{\,i\nu}{\cal
  J}_\nu)_{\vec x}\,{\cal A}^i_{\vec y}+ {\cal A}^i_{\vec
  x}\,(G^{\,i\nu}{\cal J}_\nu)_{\vec y}+ (G^{\,i-}\chi
  \,G^{\,-i})_{\vec x\,\vec y} \Bigr\rangle_{\Lambda},\,\,\,\,
 \end{equation}
where
 \begin{equation}\label{clascorrL}
  \langle{\cal A}^i_{\vec x}{\cal A}^i_{\vec y}
  \rangle_{\Lambda}\,\equiv\, \int {\cal
  D}\rho\,\,W_\Lambda[\rho]\,\,{\cal A}_a^i({\vec x}) {\cal
  A}_a^i({\vec y})\,,
 \end{equation}
with analogous formulae valid for $\langle{\cal A}^i{\cal
A}^i\rangle_{b\Lambda}$ в in terms of $W_{b\Lambda}$. Equation
(\ref{evolG}) is, in fact, an evolution equation for gluon density
that can be used for extracting the evolution equation for the
weight functional $W_\Lambda[\rho]\,\to\,W_{b\Lambda}[\rho]$.

The derivation contains two stages. First, we show that
expressions for quantum corrections can be reproduced by adding a
gaussian noise term to the r.h.s. of the classical equations of
motion (\ref{cleq0}). Second, we make corresponding redefinitions
of the classical source and weight functional.

Let us consider, therefore, the modified equations of motion
 \begin{equation}\label{cleqn}
  [D_{\nu}, F^{\nu \mu}]_a\, =\,
  \delta^{\mu+}(\rho_a(\vec x)+\nu_a(\vec x)),
 \end{equation}
now including the random source $\nu_a(\vec x)$ chosen in such a
way that on solutions of (\ref{cleqn}) the correlation $\langle
A^i A^i\rangle$ should be the same as in Theory I with quantum
corrections taken into account. The noise $\nu_a$ is thus playing
the role of charge density fluctuation $\delta\rho_a$ induced by
semihard modes. Using this analogy let us assume, that $\nu_a$ is
stationary and has the same correlators as $\delta\rho_a$:
 \begin{equation}\label{noise}\,
  \langle\nu_a(\vec x)\rangle_{\nu}\,=\,
  \sigma_a (\vec x),\qquad \langle\nu_a(\vec x)\nu_b(\vec y)\rangle_{\nu}\,=\,
  \chi_{ab}(\vec x,\vec y),
 \end{equation}
where the brackets $\langle\cdots\rangle_{\nu}$ indicate averaging
over $\nu$.

In what follows we shall need only the expansion of the solution
of (\ref{cleqn}) to the second order in noise:
 \begin{equation}\label{calAexp}
  {\cal A}^i_x[\rho+\nu]\,\approx\,{\cal A}^i_{ x}[\rho] \,+\,\frac{\delta
  {\cal A}^i_{ x}}{\delta \rho_{ y}} \bigg|_\rho\nu_{
  y}\,+\,\frac{1}{2}\, \frac{\delta^2 {\cal A}^i_{ x}} {\delta
  \rho_{ y} \delta \rho_{ z}}\bigg|_\rho\nu_{ y} \nu_{
  z}\,\equiv\,{\cal A}^i_{ x}[\rho]\,+\, \delta{ A}^i_{ x}[\rho,\nu] \,,
 \end{equation}
so that for the two-point correlator one has
 \begin{equation}\label{2PN}\,
  \langle{\cal A}^i_{ x}[\rho+\nu]\,{\cal A}^i_{y}[\rho+\nu] \rangle_\nu\,=\,
  {\cal A}^i_{ x}{\cal A}^i_{ y}\,+\,
  \langle\delta{ A}^i_{ x}\rangle_\nu {\cal A}^i_{ y}\,+\,
  {\cal A}^i_{ x}\langle\delta{ A}^i_{ y}\rangle_\nu\,+\,
  \langle\delta{ A}^i_{ x} \delta{ A}^i_{ y}\rangle_\nu\,,
 \end{equation}
where to the LLA accuracy
 \begin{eqnarray}\label{dAnn}
  \langle\delta{A}^i_{ x}\rangle_\nu&=&
  \frac{\delta {\cal A}^i_{ x}}{\delta\rho_{ y}} \bigg|_\rho\sigma(y) \,+\,
  \frac{1}{2}\, \frac{\delta^2{\cal A}^i_{ x}}
  {\delta \rho_{ y} \delta \rho_{z}}\bigg|_\rho\chi(y,z),\nn
  \langle\delta{ A}^i_{ x} \delta{A}^i_{ y}\rangle_\nu&=&
  \frac{\delta {\cal A}^i_{ x}}{\delta\rho_{ z}} \bigg|_\rho\chi(z,u)\,
  \frac{\delta {\cal A}^i_{y}}{\delta \rho_{ u}}\,.
 \end{eqnarray}
It is now easy to convince oneself that
 \begin{equation}\label{AAnu}
  \, {\cal G}(\vec x, \vec y)\,= \,
  \langle{\cal A}^i_x[\rho+\nu]\,{\cal A}^i_{ y}[\rho+\nu] \rangle_\nu\,
  \equiv\,\int {\cal D}\nu\,{\cal W}[\nu\,;\rho]\,\,
  {\cal A}^i_{ x}[\rho+\nu]\,{\cal A}^i_{ y}[\rho+\nu]\,,
 \end{equation}
where the second identity follows from averaging over $\nu$ with
the gaussian weight
 \begin{equation}\label{Wnu}\,
  {\cal W}[\nu\,;\rho]\,\equiv\,{\rm e}^{-{1\over 2}{\rm Tr}\ln\chi}
  \,\,{\rm exp}\left\{-{1\over 2}\,(\nu-\sigma)_{x}
  \chi_{x,y}^{-1}(\nu-\sigma)_{y}\right\},
 \end{equation}
(recall that $\chi$ and $\sigma$ are $\rho$ - dependent). The
correlators calculated in Theories I and II coincide provided
 \begin{equation}\label{recurW0}
  \int {\cal D}\rho\,\,W_{b\Lambda}[\rho]\,\,{\cal A}_x^i[\rho]
  {\cal A}_y^i[\rho]\,=\,\int {\cal D}\rho\,\,W_\Lambda[\rho]
  \int {\cal D}\nu\,{\cal W}[\nu\,;\rho]\,\,
  {\cal A}^i_{ x}[\rho+\nu]\,{\cal A}^i_{ y}[\rho+\nu]\,,
 \end{equation}
implying, in turn, the following recurrent relation for $W[\rho]$:
 \begin{equation}\label{recurW}
   W_{b\Lambda}[\rho]\,=\,\int {\cal D}\nu\,W_\Lambda[\rho-\nu]\,
   {\cal W}[\nu\,;\rho-\nu]\,.
 \end{equation}
Expanding the integrand of (\ref{recurW}) to second order in $\nu
\delta / \delta \rho $ and subsequently integrating over $\nu$ we
obtain
 \begin{equation}\label{EVOLW}
   W_{b\Lambda}[\rho]\,-\,W_\Lambda[\rho]\,=\,
   -\,\frac{\delta}{\delta \rho_x}\,[W_\Lambda\sigma_{x}]
   \,+\,\frac{1}{2}\,\frac{\delta^2}{\delta \rho_x \delta \rho_y}
   [W_\Lambda \chi_{xy}].
 \end{equation}
Let us stress that the convolutions in the r.h.s. of (\ref{EVOLW})
involves three-dimensional integration; e.g.
 \begin{equation}\label{rhoW1}
  \frac{\delta}{\delta \rho_x}\,[W_\Lambda\sigma_{x}] \,\equiv\,
  \int d^3{\vec x} \,\frac{\delta}{\delta \rho_a(\vec x)}\,
  \Bigl[W_\Lambda\sigma_a(\vec x)\Bigr].
 \end{equation}
Note also, that because of the support of $\sigma$ lying in the
interval $1/\Lambda^+ \simle x^- \simle 1/b\Lambda^+$, the
logarithmic enhancement appears only after integration over $x^-$.
In the limit  $b\to 1$ (\ref{rhoW1}) takes the form
 \begin{equation}\label{rhoW2}
  \frac{\delta}{\delta \rho_x}\,[W_\Lambda\sigma_{x}] \,=\,
  \alpha_s\ln{1\over b}\, \int d^2{x_\perp} \,
  \frac{\delta}{\delta \rho_a(x^-_\Lambda,x_\perp)}\,
  \Bigl[W_\Lambda\sigma_a(x_\perp)\Bigr],
 \end{equation}
where the functional derivative is calculated at the scale
$x^-_\Lambda\equiv 1/\Lambda^+$.

In terms of rapidity $\tau\equiv\ln(P^+/\Lambda^+) =\ln(1/x)$, so
that $\ln(P^+/b\Lambda^+)=\tau+\Delta \tau$ where $\Delta \tau
\equiv\ln(1/b)$. Making obvious redefinitions $W_\Lambda\equiv
W_\tau$, $W_{b\Lambda}\equiv W_{\tau+\Delta\tau}$ and
$x^-_\Lambda=1/\Lambda^+\equiv x^-_\tau$ (\ref{EVOLW}) takes the
following form
 \begin{equation}\label{DISCEVOLW}
  W_{\tau+\Delta\tau}[\rho]-W_\tau[\rho]\,=\,\alpha_s\Delta\tau
  \left\{ {1 \over 2} {\delta^2 \over {\delta \rho_\tau(x) \delta
  \rho_\tau(y)}} [W_\tau\chi_{xy}] - {\delta \over {\delta
  \rho_\tau(x)}} [W_\tau\sigma_{x}] \right\}\,,
 \end{equation}
where $\rho_\tau(x_\perp)\equiv \rho(x^-_\tau,x_\perp)$ and
convolutions are understood as two-dimensional integrals, e.g.
 \begin{equation}\label{rhoW3}
  \frac{\delta}{\delta
  \rho_\tau(x)}\,[W_\tau\sigma_{x}] \,=\, \int d^2{x_\perp}
  \,\frac{\delta}{\delta \rho_a(x^-_\tau,x_\perp)}\,
  \Bigl[W_\tau\sigma_a(x_\perp)\Bigr].
 \end{equation}
According to (\ref{DISCEVOLW}),(\ref{rhoW3}) the evolution from
$W_\tau[\rho]$ to $W_{\tau+\Delta\tau}[\rho]$ is generated by
changes in the source $\rho$ in the rapidity interval $(\tau,
\tau+\Delta\tau)$, in which the quantum corrections in the
considered LLA approximation are essential \footnote{For explicit
discussion see also \cite{AM01}. }. Note that the coordinate
support of the source is correlated with longitudinal momenta of
the modes that are integrated over. Thus rapidity $\tau$ can be
interpreted as both momentum [$\tau=\ln(P^+/\Lambda^+)$] and
coordinate [$\tau=\ln(x^-_\tau/x^-_0)$ (here $x^-_0$ -- some
arbitrary longitudinal scale, e.g. $x^-_0=1/P^+$)] rapidities.

Taking the limit of $\Delta \tau \equiv \ln(1/b)\to 0$ we arrive
at the final equation, describing the evolution of the weight
functional with $\tau\equiv\ln(1/x)\/$, first derived by direct
calculation in \cite{JKLW99a}:
 \begin{equation}\label{RGE}
  {\del W_\tau[\rho] \over {\del \tau}}\,=\,
  \alpha_s \left\{ {1 \over 2}
  {\delta^2 \over {\delta \rho_\tau(x) \delta \rho_\tau(y)}}
  [W_\tau\chi_{xy}] - {\delta \over {\delta \rho_\tau(x)}}
  [W_\tau\sigma_{x}] \right\}\,.
 \end{equation}

Equation (\ref{RGE}) is a functional Fokker-Planck equation with
$\tau$ playing the role of time, describing diffusion in the space
of color densities $\rho$ with $\rho$ -- dependent drift and
diffusion coefficients $\alpha_s\sigma$ and $\alpha_s\chi$. In the
language of probability densities (\ref{recurW}) leads to
Chapman-Kolmogorov equations. Equation (\ref{RGE}) can also be
interpreted as a functional Schroedinger equation in imaginary
time $\tau$. Evolution equation (\ref{RGE}) leads to a chain of
evolution equations on charge correlators $\langle \rho\rho
\cdots\rho\rangle_{\tau}$ \cite{JKLW99a}. For example, multiplying
$W_\tau[\rho]\,$ at $\rho(x)\rho(y)$ and performing the functional
integration over $\rho$ we get an evolution equation for the
two-point correlator
 \begin{eqnarray}\label{RGE2p}
  {d\over {d\tau}}\, \Big\langle\rho_a(\vec x)\rho_b(\vec y)\Big\rangle_\tau&=&
  \alpha_s\, \Big\langle
  \delta(x^--x^-_\tau)\sigma_a(x_\perp)\rho_b(\vec y)
  \,+\,\delta(y^--x^-_\tau)\rho_a(\vec x)\sigma_b(y_\perp)\nn
  &{}&\,\,\,\,\,\,+\,\delta(x^--x^-_\tau)\delta(y^--x^-_\tau)
  \chi_{ab}(x_\perp,y_\perp)\Bigr\rangle_\tau\,,
 \end{eqnarray}
where $\langle \cdots \rangle_\tau$ denotes averaging over $\rho$
with weight functional $W_\tau[\rho]\,$.

\subsubsection{Nonlinear evolution equation: $\alpha$ - representation}

While it is possible to compute all physical correlators in terms
of color charge densities and quantum corrections to them, it is
more illuminating to rephrase the picture of quantum evolution to
the covariant gauge and express all correlators through background
field $\alpha$ introduced in subsection \ref{classol}. The new
evolution equation reads \cite{ILM01,FILM02}:
 \begin{equation}\label{RGEa}
  {\del W_\tau[\alpha] \over {\del \tau}}\,=\,
  \alpha_s \left\{ {1 \over 2}
  {\delta^2 \over {\delta \alpha_\tau(x) \delta \alpha_\tau(y)}}
  [W_\tau\eta_{xy}] - {\delta \over {\delta \alpha_\tau(x)}}
  [W_\tau\nu_{x}] \right\}\,.
 \end{equation}
The evolution (\ref{RGEa}) includes new virtual and real kernels
$\nu$ and $\eta$. Explicit computations \cite{ILM01,FILM02} lead
to the following simple expressions for them:
 \begin{eqnarray}
  \nu^a(x_\perp) & = & {ig \over 2\pi} \int {d^2z_\perp \over
  (2\pi)^2} {1 \over (x_\perp-z_\perp)^2} {\rm Tr} \left ( T^a
  \Omega^\dagger(x_\perp) \Omega(z_\perp) \right ) \nonumber \\
  \eta^{ab}_{x_\perp,y_\perp} & = & {1 \over \pi} \int {d^2z_\perp
  \over (2\pi)^2} {(x^i-z^i)(y^i-z^i) \over (x_\perp-y_\perp)^2
  (y_\perp-z_\perp)^2} \nonumber \\
  & \times & \left \{ 1+\Omega^\dagger(x_\perp) \Omega(y_\perp)-
  \Omega^\dagger(x_\perp) \Omega(z_\perp)-
  \Omega^\dagger(z_\perp) \Omega(y_\perp) \right \}^{ab}
 \end{eqnarray}
Working with $\alpha$ - representation allows to construct a
beautiful Hamiltonian form of the evolution equation (\ref{RGEa})
first discovered by Weigert \cite{W02}:
 \begin{eqnarray}\label{RGEah}
  {\partial W_{\tau} [\alpha] \over \partial \tau} & = &
  \left \{
  \int {d^2z_\perp \over 2\pi} J^i_a(z_\perp) J^i_a(z_\perp)
  \right \} \,
  W_{\tau}[\alpha] \equiv - H W_\tau [\alpha] \nonumber \\
  J^i_a (z_\perp) & = & i \int {d^2 x_\perp \over 2\pi}
  {z^i-x^i \over (z_\perp-x_\perp)^2}
  \left(
  1-\Omega^\dagger(z_\perp) \Omega(x_\perp)
  \right ) _{ab}
  {\delta \over \delta \alpha_\tau^b(x_\perp)}
 \end{eqnarray}
As mentioned above, evolution equation (\ref{RGEa}) allows to
calculate arbitrary correlators of $\alpha$ - dependent vertices.
The evolution equation for one combination of special interest,
namely ${\cal V} (x_\perp, y_\perp) \equiv {\rm tr} \left(
\Omega^\dagger(x_\perp) \Omega (y_\perp) \right ) $ reads
\cite{FILM02}
 \begin{eqnarray}\label{B}
  {\partial {\cal V} (x_\perp,y_\perp) \over \partial \tau} & = &
  - {\alpha_s \over 2\pi^2} \int d^2z_{\perp} {(x_\perp-y_\perp)^2
  \over (x_\perp-z_\perp)^2(y_\perp-z_\perp)^2 } \nonumber \\
   & \times & \langle N_c {\cal V} (x_\perp, y_\perp) -
  {\cal V} (x_\perp, z_\perp) {\cal V} (z_\perp, y_\perp) \rangle _{\tau}
 \end{eqnarray}
This equation was originally derived by Balitsky \cite{B96} using
a formalism of functional operator product expansion on the light
cone.

\subsubsection{Nonlinear evolution equation: results}

Before turning to describing the known (approximate) analytical
solutions of (\ref{RGEa}) let us discuss a simple truncation of
the hierarchy of equations (\ref{B}) that reduces all higher order
correlators to products of the basic two-point function, e.g.
 $$
  \langle {\cal V} (x_\perp, z_\perp) {\cal V} (z_\perp, y_\perp) \rangle _{\tau} \to
  \langle  {\cal V} (x_\perp, z_\perp) \rangle _{\tau}
  \langle  {\cal V} (z_\perp, y_\perp) \rangle _{\tau}
 $$
One way to achieve such factorization "automatically" is to work
in the large - $N_c$ limit. The above-defined quantity ${\cal V}
(x_\perp, y_\perp)$ is related to the scattering amplitude ${\cal
N} (r_\perp \equiv x_\perp-y_\perp)$ of a corresponding (dependent
on the representation of gauge group used in constructing the
Wilson lines) color dipole via
 \begin{equation}
  {\cal N} (r_\perp) \, = \, {1 \over N_c} \left ( {\rm tr(1)} - {\cal V} (x_\perp,y_\perp) \right ).
 \end{equation}
Corresponding evolution equation for ${\cal N}$, first derived by
Balitsky \cite{B96}, reads
 $$
  {\partial {\cal N} (x_\perp,y_\perp) \over \partial \tau} \, = \,
  - {\alpha_s \over \pi} \int d^2z_{\perp} {(x_\perp-y_\perp)^2
  \over (x_\perp-z_\perp)^2(y_\perp-z_\perp)^2 }
 $$
 \begin{equation}\label{BK}
    \times  \left [ {\cal N} (x_\perp, z_\perp) + {\cal N} (z_\perp, y_\perp) - {\cal N} (x_\perp, y_\perp)
  - {\cal N} (x_\perp, z_\perp) {\cal N} (z_\perp, y_\perp) \right ] _{\tau}
 \end{equation}
Equation (\ref{BK}) was independently derived by Kovchegov
\cite{K9900} in the framework of color dipole model formalism and
later rederived in \cite{BR} by direct summation of fan diagrams.
It is important to note that ${\cal N} (x_\perp, y_\perp)$ is, in
fact, a scattering amplitude {\it at fixed impact parameter}
$b_\perp = (x_\perp+y_\perp)/2$, ${\cal N} (x_\perp, y_\perp)
\equiv {\cal N}(b_\perp,r_\perp)$, where $r_\perp =
x_\perp-y_\perp$. This detail will be of importance in the next
subsection in discussing the interrelation between saturation and
unitarity.

The properties of the solution of (\ref{BK}) are best illustrated
by its convenient parametrization introduced in \cite{GW99}
 \begin{equation}\label{amsat}
  {\cal N} (x_\perp,y_\perp) \, = \, 1 - {\rm exp} \left [-(r_\perp)^2 Q^2_s(\tau,b_\perp) \right ].
 \end{equation}
In equation (\ref{amsat}) the energy dependence of the scattering
amplitude ${\cal N}$ is controlled by the $\tau$- dependence of
the saturation momentum $Q^2_s$. Extensive numerical and
theoretical analysis \cite{AM99,GW99,IM01,LT0001} showed that the
following simple parametrization of the energy dependence of the
saturation momentum $Q_s (b_\perp,\tau)$ is valid:
 \begin{equation}\label{modsatmom}
  Q^2_s (b_\perp,\tau) \, = \,  Q^2_s (b_\perp,\tau_0) {\rm e}^{\lambda \alpha_s (\tau-\tau_0)},
 \end{equation}
where $\lambda$ is the numerical coefficient, $\lambda \sim 1$.
From equations (\ref{amsat}),(\ref{modsatmom}) we see that the
magnitude of the scattering amplitude is determined by the
multiplicative combination of the probe size $Q_\perp^2 \sim
1/r_\perp^2$ and rapidity $\tau$. At large $Q^2_\perp$ and
moderate $\tau$ we have a usual perturbative answer ${\cal N} \sim
1/Q^2_\perp$. Most interesting is, of course, the high energy
limit $\tau \to \infty$, in which, due to ${\bar Q^2}_s (\tau \to
\infty) \to \infty$ (cf. (\ref{modsatmom})), the second term in
(\ref{amsat}) vanishes, and the scattering probability saturates
at its upper border, ${\cal N} (\tau \to \infty) \to 1$. Thus, the
quadratic nonlinearity in the kernel of Balitsky-Kovchegov
equation (\ref{BK}) ensures unitarization {\it at fixed impact
parameter} $b_\perp$. Transition between the purely perturbative
medium energy regime to the nonlinear high energy one is
controlled by the key object of the theory - saturation momentum.

Discussing the solutions of the nonlinear renormalization group
equation (RGE) (\ref{RGEa}) or its hamiltonian counterpart
(\ref{RGEah}), it is necessary to take a closer look at the
coordinate structure of functional derivatives over $\alpha$ in
these equations \cite{FILM02,IM01}, see also \cite{AM01a}. The
main point is that, as follows from the experience obtained in
computing quantum corrections to MV model, the quantum evolution
developed up to the scale $\tau$ generates a field $\alpha$ having
support in the interval $0 \leq x^- \leq {\rm e}^{\tau}/P^+ \equiv
x_0^- {\rm e}^{\tau} \equiv x^-_{tau}$. Then, within the
constructed RG procedure, one can make the following replacement
 \begin{equation}\label{rmm}
  \Omega^\dagger (x_\perp) \to \Omega^\dagger_{\tau} (x_\perp)
  \equiv {\rm P} \exp \left \{ ig \int_0^{x^-_{\tau}} dx^-
  \alpha(x^-,x_\perp) \right \}.
 \end{equation}
Quantum evolution takes place at the border of the covered
kinematical interval, so the functional derivatives of the Wilson
lines (\ref{rmm}) in the RGE (\ref{RGEa},\ref{RGEah}) are in fact
taken with respect to the color field $\alpha_{\tau}(x_\perp)
\equiv \alpha(x^-_\tau,x_\perp)$ at the end point $x^-_{\tau}$:
 \begin{equation}\label{fd}
  {\delta \Omega^\dagger_{\tau} (x_\perp) \over \delta \alpha^a_{\tau}
  (z_\perp)} \, = \, ig\delta^{(2)}(x_\perp - x_\perp) T^a \Omega^\dagger_{\tau} (x_\perp)
 \end{equation}
This detail is very important in discussing the general properties
of the solution of the master equation (\ref{RGE}),(\ref{RGEa})
and its physical interpretation \cite{IM01,IIM02a,IIM02b}. It was
proven, in particular, that for very different reasons both small
transverse momentum ($q^2_\perp \ll Q^2_s$) and large transverse
momentum ($q^2_\perp \gg Q^2_s$) asymptotics are described by the
gaussian weight functional $W$ satisfying
 \begin{equation}\label{RGEg}
   {\partial W_{\tau} [\rho] \over \partial \tau} \, = \, {1 \over 2}\
   \int_{x_\perp,y_\perp} \lambda(x_\perp,y_\perp)
   {\delta^2 W_{\tau} [\rho] \over \delta \rho^a_{\tau} (x_\perp) \delta \rho^a_{\tau} (y_\perp)}.
 \end{equation}
At large transverse momenta one can simply neglect all nonlinear
effects, so a complete description of the system is given by the
two-point function solving the BFKL equation. At small transverse
momenta the situation is again gaussian, but this time due to the
fact that nonlinearities in the Wilson lines forming the kernel
become vanishingly small because of the rapid oscillations of the
Wilson lines. It turned out possible to prove that, to a good
accuracy, the solution of the generic master equation
(\ref{RGE}),(\ref{RGEa}) can be approximated by that of its
gaussian counterpart (\ref{RGEg}) having the following kernel,
interpolating between the small transverse momentum and large
transverse momentum regimes:
 \begin{equation}\label{gkern}
  \lambda (k_\perp) \, = \, \lambda^{BFKL}_{\tau} \cdot {k^2_\perp \over k^2_\perp + \pi \lambda^{BFKL}_{\tau}}
 \end{equation}

Let us also mention one more important observation made in
\cite{IM01,IIM02a}, namely, the "geometric scaling" behavior of
the kernel
 \begin{equation}\label{gs}
  \lambda_{\tau}(k_\perp) \simeq {1 \over \pi} k_\perp^2 \left ( Q^2_s(\tau) \over k^2_\perp \right )^\gamma
 \end{equation}
valid within the "scaling window"  of  $Q^2_s(\tau) \ll k^2_\perp
\ll Q^4_s(\tau)/Q^2_0$ and ensuring the corresponding geometric
scaling behavior of physical observables.

\subsection{Saturation and unitarity}

Equation (\ref{amsat}) demonstrates the saturation phenomenon in
terms of the dipole scattering amplitude. Indeed, it is obvious
from Eq.~(\ref{amsat}) that the scattering amplitude for large
size dipoles (those with $r_\perp^2 \equiv (x_\perp-y_\perp)^2 >
Q^2_s$ ) is vanishingly small. It is important to note that the
r.h.s. of Eq.~(\ref{amsat}) depends only on the size of the dipole
$r_\perp \equiv x_\perp-y_\perp$ and thus holds for any {\it
given} impact parameter $b \equiv (x_\perp+y_\perp)/2$. Unitarity
(i.e. condition $N <= 1$) is therefore ensured only at given $b$.

In the previous paragraph we have seen that saturation phenomenon
restores unitarity at given impact parameter. Most important
question to answer at this stage is thus whether saturation helps
to solve the unitarity violation problem for the total inelastic
cross section, obtained by integrating the scattering amplitude
over the impact parameter, as well. Unfortunately, the answer to
this question turns out to be negative. A detailed analysis of
this problem can be found in \cite{KW1,KW2,KW3,FIIM02}.

The unitarity requirement leads to the famous Froissart bound on
the maximal allowable growth of the total inelastic cross-section
with energy
\begin{equation}\label{Froissart}
 \sigma_{inel} < {\pi \over m^2}\,\tau^2,
\end{equation}
where $m$ is the smallest mass scale in the theory (pion mass for
QCD with light quarks).

The physical cross-section for the probe having the transverse
size $Q^2_\perp \sim 1/r^2_\perp$ at energy $s \sim \exp (\tau)$
is obtained by integrating the scattering probability ${\cal N}$
over the impact parameter $b_\perp$:
 \begin{equation}\label{uncross}
  \sigma (Q^2_\perp, \tau) \, = \, 2 \int d^2 b_\perp {\cal N} (Q_\perp,b_\perp | \tau)
  \equiv \pi R^2 (\tau)
 \end{equation}
where we have introduced an energy-dependent interaction radius
$R(\tau)$. Expressed in terms of this radius, the Froissart bound
Eq.~(\ref{Froissart}) corresponds to the maximal possible growth
of the radius of interaction radius of $R(\tau) \sim \tau$.

The key reason for unitarity breakdown is easily understood by
noting, that in perturbation theory the decay of the fields at
infinity, be it a color singlet or open color system, is always
powerlike. So, at large enough impact parameters $b_\perp$ and
high enough energies the integral in (\ref{uncross}) diverges
exponentially. This effect can be summarized by a compact formula
for the total cross-section derived in \cite{KW1}:
 \begin{equation}\label{univio}
  \sigma_{inel} \, = \, \pi R^2_{target} + 2 \pi R_{target} x_0\exp
  \left [ {\alpha_s N_c \over 2 \pi} \epsilon \tau \right ],
 \end{equation}
where $\epsilon$ is a constant. From Eq.~(\ref{univio}) we see,
that for $\tau > 1/(\epsilon \alpha_s) \ln R_{target}/x_0$ the
divergence of the total cross-section is exponential and thus
violates the Froissart bound. Another way of understanding the
reason for unitarity violation is to observe that no perturbative
mass transmutation is possible in the massless non-abelian theory,
so the perturbation theory can not generate within itself a mass
scale $m$ (cf. Eq.~(\ref{Froissart})) that can convert the
powerlike dependence on impact parameter into the exponential one
and save unitarity. Thus, there is still a lot of important
nonperturbative physics beyond the nonlinear effects grasped by
the improved perturbation theory (for a transparent discussion in
terms of constituent quarks, soft pomeron, etc., see \cite{KW2}).


\section{Dense gluon matter in nuclear collisions}

Of exceptional interest to the studies of nonabelian parton
dynamics are ultrarelativistic heavy ion collisions. They provide
dense initial partonic fluxes and thus conditions for creation of
dense partonic matter at the early stages of collision.

To describe the parton - related dynamics in high energy nuclear
collisions it is necessary to specify the role of partonic degrees
of freedom within the chosen description of nuclear interaction.
Below we will discuss the two approaches used in the literature
\footnote{It is, of course, possible to analyze the dynamics of
nuclear collisions in terms of qluon strings stretched between
constituent degrees of freedom, see e.g. \cite{KM84}.}.

In the first one, discussed in subsection \ref{mixed},
 one introduces \cite{EKL,HIJING} a mixture of soft
nonperturbative (hadronic strings, etc.) and semihard perturbative
contributions to the inelastic cross-section. The perturbative
cross-sections are strongly divergent at small momentum transfer
(small transverse energy), so to arrange a finite contribution to
the inelastic cross-section one has to introduce an explicit
infrared cutoff. The dominant contribution to perturbative
component of inelastic cross section will, therefore, come from
the infrared cutoff scale - a situation that is not conceptually
satisfactory (the infrared divergence of the perturbative
cross-sections is typically powerlike, so introducing a cutoff
presents a "brutal" way of fixing a physically important scale).

In the second approach \cite{BM,AM00a}, described  in subsection
\ref{sat}, and related to physical implications of gluon
saturation phenomenon, the appearance of characteristic momentum
scale, the saturation scale $Q_s$, is not artificial, but is a
natural consequence of the nonlinear effects in dense gluon
medium. Technically speaking, the saturation momentum $Q_s$ also
plays a role of the infrared regulator, and the dominant
contribution to physical cross-section is again coming from the
vicinity of this momentum scale. A crucial difference with respect
to the mixed models is that in saturation type models it is
extremely difficult, if possible, to consistently add soft
nonperturbative component.

\subsection{Mixed model}\label{mixed}

The most complete description of the dynamics of heavy ion
collision based on superposition of soft stringy and semihard
partonic dynamics currently available is provided by HIJING model
\cite{HIJING}. In addition to taking into account initial and
final state radiation in hadronic collisions, the model also
accounts for nuclear shadowing of the structure functions and the
energy loss of produced partons in the debris created in nuclear
collision. The nuclear collision is described as a superposition
of nucleon-nucleon ones. The $pp$ block of HIJING was "normalized"
on experimental data at energies $\sim 100$ GeV, so the model is
very well tuned to RHIC energies. Although it is probably very
difficult to extrapolate, without major changes, the physics
embedded into HIJING to LHC energies, at RHIC energies the model
provides a relatively consistent and reliable framework for
analyzing the physics of nuclear collisions.

A good illustration of the usage of mixed "soft-hard" approach is
provided by the discussion of the energy and centrality dependence
of the (pseudo)rapidity particle density \cite{GW01,KN01}. The
formula for this density includes a typical mixture of soft and
hard contributions:
\begin{equation}\label{msh}
 {dN \over dy} \, = \, (1-X(s))\ n_{pp}\
 { \langle N_{part} \rangle \over 2} +
 X(s)\ n_{pp}\ \langle N_{coll} \rangle
\end{equation}
where $n_{pp}$ is a pseudorapidity particle density in $pp$
collisions, $X(s)$ is a yield of semihard dynamics in particle
production and $\langle N_{part(coll)} \rangle$ are the average
numbers of participants (collisions) at given energy. Let us
remind that by "participants" one understands counting nucleons
experiencing at least one inelastic collisions, while the
subscript "coll" refers to counting all inelastic collisions.  The
estimate of $X(s)$ in \cite{KN01} gave $X(130\ {\rm GeV}) \sim
0.1$, i.e. a ten percent share of semihard particle production
mechanisms.


\subsubsection{Anatomy of the transverse energy flow}

Before turning to nuclear collisions, we discuss \cite{L00} in
this subsection the "anatomy" of the transverse energy flow
generated in multiparticle production processes in nucleon-nucleon
interactions in terms of the relative contributions of various
perturbative and nonperturbative mechanisms. As mentioned before,
the nucleon-nucleon collision is a basic building block in the
mixed type models like HIJING, so the results of this analysis
will help to gauge, through comparison with experimental data
\cite{UA2}, the role of different mechanisms contributing to the
observed transverse energy flow and their physical interpretation.

Let us first turn to the calculation of the perturbative
contribution to the transverse energy flow in the central rapidity
window in the next-to-leading (NLO) order accuracy and compare it
to experimental data obtained by UA2 collaboration \cite{UA2}. The
NLO calculation of a generic jet cross section requires using a
so-called jet defining algorithm specifying the resolution for the
jet to be observed, for example, the angular size of the
jet-defining cone, see e.g. \cite{S}. The cross section in
question is calculated by integrating the differential one  over
the phase space, with the integration domain restricted by the jet
characteristics  fixed by the jet-defining algorithm. The NLO
distribution of the transverse energy produced into a given
rapidity interval $y_a < y < y_b$ is given, to the $O(\alpha_s^3)$
order,  by
\begin{eqnarray}\label{nlojet}
 \frac{d\sigma}{dE_\perp}=\int D^2PS  \frac{d\sigma}{d^4p_1d^4p_2}
 \delta(E_\perp-\sum\limits_{i=1}^2 |p_{\perp
 i}|\theta(y_{min}<y_i<y_{max}))\nonumber\\
 +\int D^3PS  \frac{d\sigma}{d^4p_1d^4p_2d^4p_3}
 \delta(E_\perp-\sum\limits_{i=1}^3 |p_{\perp
 i}|\theta(y_{min}<y_i<y_{max})),
\end{eqnarray}
where the first contribution corresponds to the two-particle final
state with one-loop virtual corrections taken into account and the
second contribution  to the three-particle state.

 In perturbative QCD one can rigorously compute only infrared safe
quantities \cite{S}, in which the divergences originating from
real and virtual gluon contributions cancel each other, so that
adding very soft gluon does not change the answer. It is easy to
convince oneself, that the transverse energy distribution into a
given rapidity interval Eq.~(1) satisfies the above
requirement\footnote{For a formal definition of infrared safety
see, e.g., \cite{KS}.}.

 The calculation  of  transverse energy spectrum in $p {\bar p}$ collisions
was  performed in \cite{LO01} using the Monte-Carlo code developed
by Kunzst and Soper \cite{KS}, and a  "jet" definition appropriate
for transverse energy production Eq.~(\ref{nlojet}).

\begin{figure}[h]
 \begin{center}
 \epsfig{file=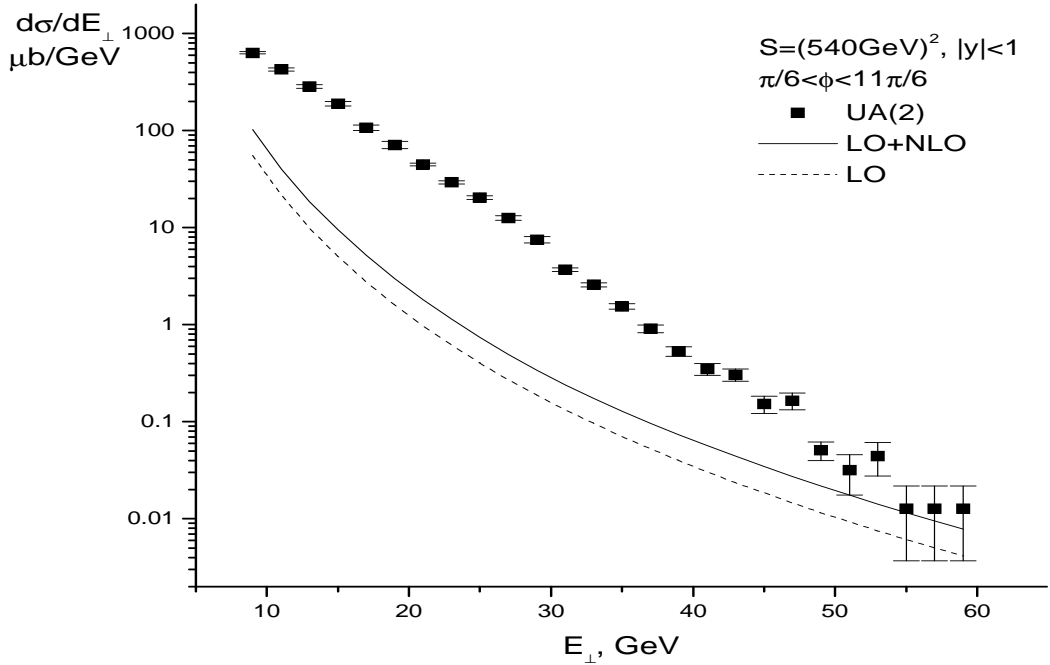,height=9cm,width=14cm}
 \end{center}
 \caption{Transverse energy spectrum in $p {\bar p}$ collisions calculated
  within LO+NLO accuracy in perturbative QCD vs the experimental data by
  UA2 collaboration \cite{UA2}.}
 \label{fua2}
\end{figure}

In Fig.~\ref{fua2} we compare the  LO and LO+NLO transverse energy
spectra for $p{\bar p}$ collisions, calculated following the
procedure described in \cite{LO99}, with the experimental data on
transverse energy distribution in the central rapidity window $|y|
< 1$ and azimuthal coverage $\pi/6 \le \varphi \le 11 \pi/6$ at
$\sqrt{s}=540$ GeV measured by UA2 Collaboration \cite{UA2}.
We see that the perturbative LO+NLO calculations start merging
with the experimental data only around quite a large scale
$E_{\perp} \sim 60$\ GeV. It is interesting to note, that it is
precisely around this energy, that the space of experimental
events starts to be dominated by two-jet configurations
\cite{UA2}. This means that only starting from these transverse
energies the assumptions behind the perturbative calculation
(collinear factorization at leading twist, explicit account for
all contributions of a given order in $\alpha_s$) are becoming
adequate to the observed physical process of  transverse energy
production providing the required duality between the  description
of  dominant configuration contributing to transverse energy
production at this order in perturbation theory and the final
state transverse energy carried by hadrons. At $E_{\perp} \le 50$\
GeV the calculated spectrum is in radical disagreement with the
experimental one both in shape and magnitude calculated and
observed spectrum is very large indicating the inadequacy of the
considered $O(\alpha_s^3)$  perturbative calculation in this
domain. Let us mention here, that it is currently impossible to
improve the results of the above calculation, because neither
calculations of higher order nor infinite order resummation for
this process are currently available.

\begin{figure}[h]
 \begin{center}
 \epsfig{file=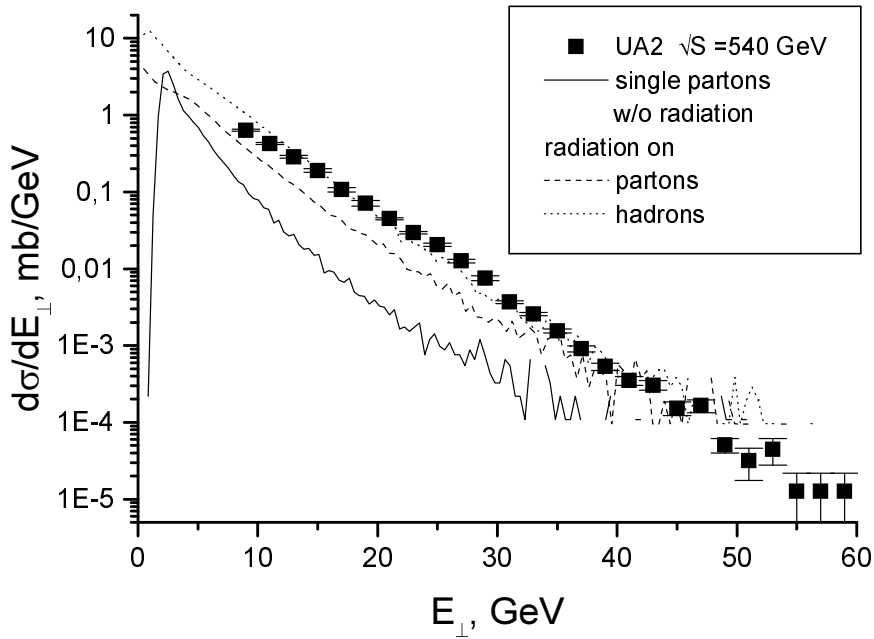,height=9cm,width=12cm}
 \end{center}
 \caption{Transverse energy spectrum in $p {\bar p}$ collisions
 calculated in HIJING vs the experimental data by UA2 collaboration
 \cite{UA2}.}
 \label{fua2hj}
\end{figure}

In practical terms this means that additional model assumptions
are needed to achieve agreement with experimental data  strongly
indicating that higher order corrections and higher twist effects
have to be taken into account (in a necessarily model-dependent
way) in order to describe them. In the popular Monte-Carlo
generators such as PYTHIA \cite{PYTHIA} and HIJING \cite{HIJING}
such effects as multiple binary parton-parton collisions, initial
and final state radiation and transverse energy production during
hadronization are included. In Fig.~\ref{fua2hj} we compare the
same experimental data by UA(2) \cite{UA2} with the spectrum
calculated with HIJING event generator. To show the relative
importance of different dynamical mechanisms, in Fig.~\ref{fua2hj}
we plot the contributions from the hard parton scattering without
initial and final state radiation, full partonic contribution and,
finally, the transverse energy spectrum of final hadrons.
We see, that taking into account additional partonic sources such
as, e.g.,  initial and final state radiation, allows to reproduce
the (exponential) form of the spectrum, but still not the
magnitude. The remaining gap is filled in by soft contributions
due to transverse energy production from decaying stretched
hadronic strings. Let us finally note, that the spectrum
calculated in HIJING is somewhat steeper than the experimental
one. Additional fine-tuning can be achieved by probing different
structure functions.

The above results clearly demonstrate that in order to reproduce
the experimentally observed transverse energy spectrum, one has to
account for complicated mechanisms of parton production, such as
initial and final state radiation accompanying hard parton-parton
scattering, production of gluonic kinks by strings, as well as for
nonperturbative transverse energy production at hadronization
stage. This statement is a calorimetric analog of the well-known
importance of the minijet component in describing the tails of the
multiplicity distributions, \cite{SZ} and \cite{HIJING}.

Let us note that the result has straightforward implications for
describing the early stages of heavy ion collisions. In most of
the existing dynamical models of nucleus-nucleus collisions they
are described as an incoherent superposition of nucleon-nucleon
ones. As we have seen, to correctly describe the partonic
configuration underlying the observed transverse energy flow in
nucleon-nucleon collisions, mechanisms  beyond conventional
collinear factorization are necessary. This means that  to
estimate such quantities as, for example, parton multiplicity at
some given timescale, a very careful analysis of various
contributions is required.


\subsubsection{Azimuthal pattern of transverse energy flow}

Understanding the parton-related dynamical features of heavy ion
collision calls for the analysis of experimentally observable
quantities sensitive to particular features, distinguishing
semihard parton dynamics from the soft hadronic one. One specific
proposal in this direction was discussed in \cite{LO00,LO01}. The
idea is that perturbative energy production generates asymmetric
transverse energy flow due to its collimation along the directions
fixed in the process of large momentum transfer.

To quantify the  event-by-event asymmetry of transverse energy
flow, let us consider the difference between the transverse energy
deposited, in some rapidity window $y_{min}<y_i<y_{max}$, into two
oppositely azimuthally oriented sectors with a specified angular
opening $\delta \varphi$ each.

\begin{figure}
\vspace*{-10mm}
\begin{center}
 \epsfig{file=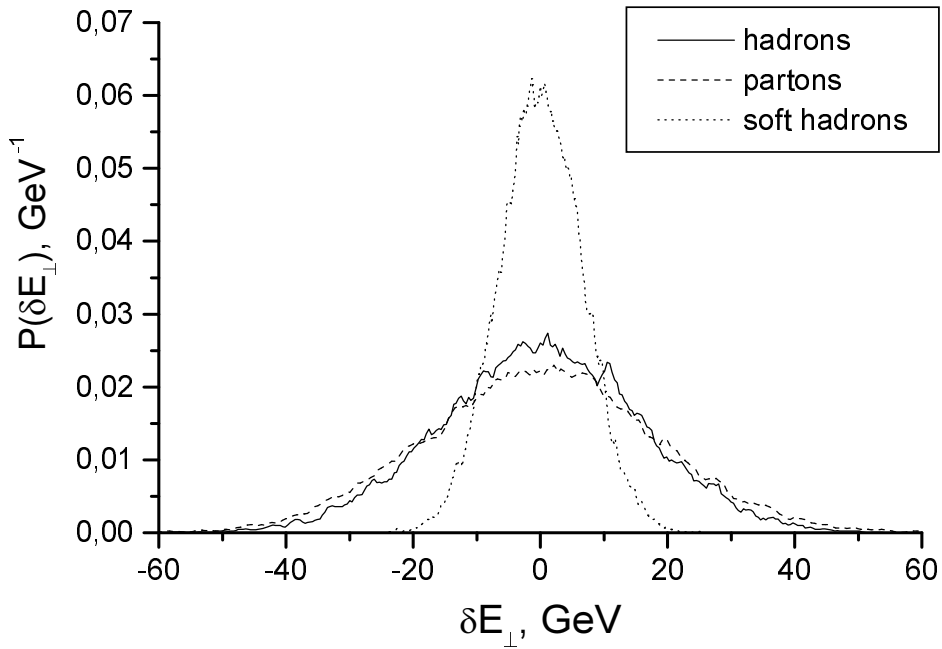,height=9cm,width=13cm}
 \end{center}
\vspace*{-5mm}
 \caption{Probability distribution for azimuthal transverse
energy disbalance in the unit rapidity window for AuAu collisions
at RHIC energy $\sqrt{s}=200\,{\rm GeV}$, $p_0=2\,{\rm GeV}$,
quenching on.}
 \label{frhic}
\end{figure}

\begin{figure}
\vspace*{-5mm}
 \begin{center}
 \epsfig{file=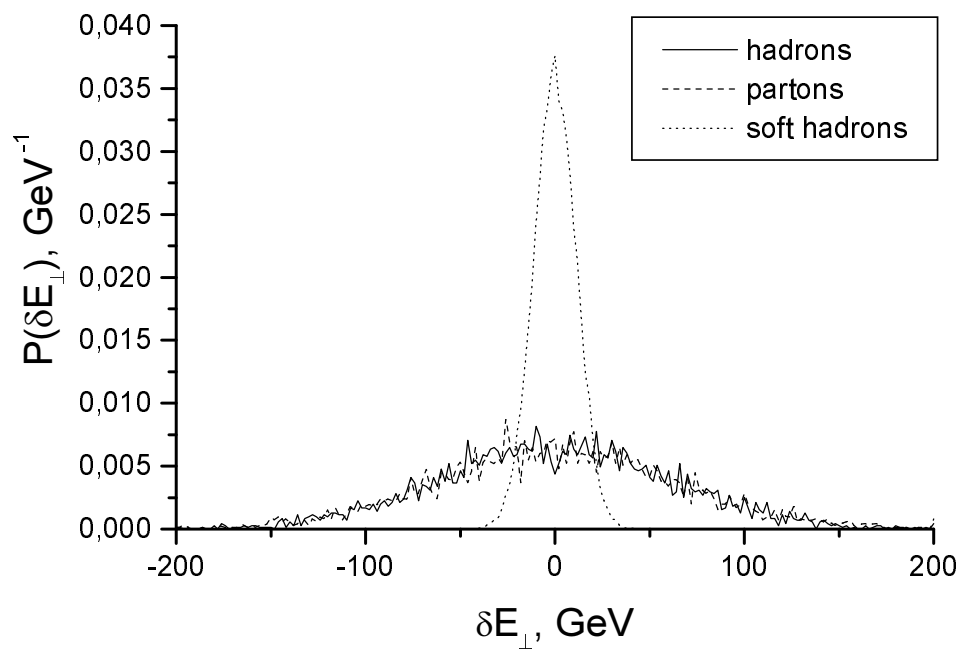,height=9cm,width=13cm}
 \end{center}
\vspace*{-5mm}
 \caption{Probability distribution for azimuthal transverse
energy disbalance in the unit rapidity window for PbPb collisions
at LHC energy $\sqrt{s}=5.5\,{\rm TeV}$, $p_0=2\,{\rm GeV}$,
quenching on. }
 \label{flhc}
\end{figure}

For convenience one can think of the directions of these cones as
being  "up" and "down" corresponding to some specific choice of
the orientation of the system of coordinates in the transverse
plane. All results are, of course, insensitive to the particular
choice. Denoting now the transverse energy going into the "upper"
and "lower" cones in a given event by $E_{\perp}^{\uparrow}
(\delta \varphi)$ and $E_{\perp}^{\downarrow}(\delta \varphi)$
correspondingly, we can quantify the magnitude of the asymmetry in
transverse energy production in a given event by \be\label{deltaE}
 \delta E_{\perp} (\delta \varphi) \, = \,
 E_{\perp}^{\uparrow} (\delta \varphi) -
 E_{\perp}^{\downarrow} (\delta \varphi),
\ee its statistical properties characterized by the corresponding
probability distribution \be\label{distr}
  P(\delta E_{\perp}|{\delta \varphi}) \, = \,
 {d\,w(\delta E_{\perp} (\delta \varphi)) \over d\,\delta E_{\perp}
 (\delta \varphi)}.
\ee
 This distribution was calculated \cite{LO01}, in HIJING model,
for central AuAu collisions at RHIC energy $\sqrt{s}=200\,{\rm
GeV}$ and central PbPb collisions at LHC energy
$\sqrt{s}=5.5\,{\rm TeV}$ for $\delta \varphi=\pi$. The
distributions $P(\delta E_\perp|\pi)$ have been calculated both at
partonic level and at the level of final hadrons with semihard
interactions and quenching on and off. This allowed us to study
the contribution of HIJING minijets and of the effects of their
hadronization to the asymmetry in question. The resulting
distributions are plotted in
 Figs.~\ref{frhic} and \ref{flhc}, for RHIC and LHC energies respectively
with quenching turned on and the value of the minijet's infrared
cutoff $p_0=2\,{\rm GeV}$.

The numerical values of the mean square deviation $\delta
E_{\perp}$ characterizing the widths of the corresponding
probability distributions in Figs.~\ref{frhic} and \ref{flhc} are
given in Table 1, where for completeness we also give the widths
for the probability distributions with quenching turned off and
with a larger value for the infrared cutoff $p_0=4\,{\rm GeV}$.
\begin{center}
\begin{tabular}{|c|c|c|l|c|}
\hline AA&$\sqrt{S}$, GeV&$p_0$, GeV & asymmetry &
$\sqrt{\left<\delta E^2\right>}$\strut\\
\hline
            &            &         & hadrons (quenching on)     & 16\\
\cline{4-5}
\halfs{AuAu}&\halfs{200} &\halfs{2}& hadrons (quenching off)    & 17\\
\cline{4-5}
            &            &         & partons                    & 18\\
\cline{4-5}
            &            &         & soft hadrons               &  7\\
\hline
            &            &         & hadrons (quenching on)     & 61\\
\cline{4-5}
\halfs{PbPb}&\halfs{5500}&\halfs{2}& hadrons (quenching off)    & 71\\
\cline{4-5}
            &            &         & partons                    & 65\\
\cline{4-5}
            &            &         & soft hadrons               & 15\\
\hline
            &            &         & hadrons (quenching on)     & 69\\
\cline{4-5}
PbPb        & 5500       &        4& partons                    & 76\\
\cline{4-5}
            &            &         & soft hadrons               & 16\\
\hline
\end{tabular}\nopagebreak

\bigskip
{\bf Table 1}
\end{center}

 The main conclusions that can be drawn from Figs.~\ref{frhic} and
\ref{flhc} and Table 1 are the following.

 First, the magnitude of the azimuthal asymmetry as measured
by the width of the probability distribution $P(\delta
E_{\perp}|\delta \varphi)= dw(\delta E_{\perp} (\delta
\varphi))/d\delta E_{\perp}(\delta \varphi)$ is essentially
sensitive to semihard interactions (minijets).
 Switching off minijets, and thus restricting oneself
to purely soft mechanisms, leads to a substantial narrowing of the
asymmetry distribution; by the factor of 2.3 at RHIC and by the
factor 4.1  at LHC energy respectively (these values correspond to
the case of quenching being turned on).

Second, quite remarkably, the parton and final (hadronic)
distributions of $\delta E_{\perp}$ in both cases practically
coincide indicating that the contribution to transverse energy due
to hadronization of the initial parton system is, with a high
accuracy,  additive and symmetric in between the oppositely
oriented cones. Both  conclusions show that the energy-energy
correlation in Eq.~(\ref{deltaE})  is a sensitive  measure of the
primordial parton dynamics that can be studied in calorimetric
measurements in central detectors at RHIC and LHC.

Third, as expected, turning off quenching somewhat enhances the
fluctuations. However, as seen from the table, numerically the
effect is not important. This shows once again that the proposed
asymmetry is really essentially determined by the earliest stage
of the collision, when the primordial parton flux is formed.

Finally, from Table 1 we conclude that the studied asymmetry is
not particularly sensitive to changing the value of the infrared
cutoff $p_0$ and thus provides a robust signal for the presence of
semihard dynamics deserving an experimental study.


\subsubsection{Turbulent initial glue: impact parameter plane
picture}\label{turb}

In the previous paragraph we have discussed an event-by-event
asymmetry of the transverse energy flow from the "momentum" point
of view. In a more detailed analysis a spatial pattern giving rise
to this energy-momentum flux should be considered. Of particular
interest is an event-by-event transverse energy generation pattern
in the impact parameter plane. This question was first addressed
in \cite{GRZ97} - with strikingly interesting results.

The event-by-event transverse energy release pattern in the
mixed-type models like HIJING is determined by two major factors.
The first one is a distribution of the number of soft and
(semi)hard inelastic collisions per unit transverse area. The
second is the shape of the corresponding transverse momentum
(energy) spectra. The convolution of the two distributions
determines a shape of the transverse momentum and energy release.
For broad resulting distributions one expects an intermittent
turbulent-like spatial transverse energy distribution in the
transverse plane. Most promising in this respect is of course the
semihard partonic component. The distribution in the number of
semihard inelastic interactions is quite broad, and the transverse
energy spectra generated in these collisions are powerlike. It is
precisely this combination that leads to the intermittent
turbulent-like pattern of the primordial transverse energy release
\cite{GRZ97}.

In \cite{GRZ97} the transverse energy distribution for an ensemble
of free-streaming gluons (taken from the HIJING parton event list)
at zero rapidity $y=0$ (and thus at $z=0$) at proper time $\tau$
was taken to be
\begin{equation}\label{turben}
    {\cal E} (\tau,x_{\perp},z=0) \, = \, \sum_k {p_{\perp \, k} \over \tau}\,
    {(\tau p_{\perp \, k})^2 \over 1+(\tau p_{\perp \, k})^2} \,
    \delta (x_{\perp}-x_{\perp \, k} (\tau)) \, \delta (y_k),
\end{equation}
where summation is over partons and the second factor in the
right-hand side stands for the parton formation probability
distribution.

To be meaningful, the transverse energy distribution should be
considered in the coarse-grained coordinate space. More
particularly, the size of the transverse cell is actually limited
from below by the uncertainty principle $\delta r_{\perp} > 1 /
\delta p_{\perp}$ and from above by causality (local horizon of
the gluon in the comoving frame). At given $\tau$ this upper bound
is simply given by $\delta r_{\perp} < \tau$, so that for large
nuclei and small proper times the number of independent cells in
the transverse plane can be quite large. The natural longitudinal
size of the cell can be chosen as $|y|<1$.

The particular case considered in \cite{GRZ97} was Au-Au
collisions at RHIC energy $\sqrt{s}=200\,\, {\rm GeV}$. The
snapshot of the transverse energy and transverse momentum
distribution was taken at $\tau = 0.5\,\, {\rm fm}$. The results
turned out to be quite striking. On the background of the smooth
uniformly distributed energy density ${\cal E}_{soft} \simeq 5\,\,
{\rm GeV}$ one finds pronounced peaks ("hot spots") with large
energy densities ${\cal E} > 20\,\, {\rm GeV}$ (corresponding to
${\cal E}_{hard} \succeq 15\,\, {\rm GeV}$) separated by distances
of order $4-5\,\, {\rm fm}$, and the momentum field showed
nontrivial vortex-like structure. A natural analogy suggested in
\cite{GRZ97} was that of the instabilities (turbulence) induced in
the uniform "soft" laminar flow by the minijet component.

The importance of the results of \cite{GRZ97} go, in our opinion,
far beyond the particular model (HIJING), collision energy, etc.,
considered in the paper. As we have mentioned on many occasions in
the preceeding paragraphs, a consistent model of heavy ion
collisions is necessarily a mix of soft and hard mechanisms. Any
such mix will generate a turbulent-like intermittent picture
analogous to the one discussed in \cite{GRZ97}.


\subsection{Parton production and saturation in nuclear
collisions}\label{sat}

The cross-sections describing hard processes (e.g. high-$p_\perp$
jet production) are proportional to the product of incoming
partonic flows. At high energies, when the saturation phenomenon
becomes important, this customary picture has to be reconsidered.
This analysis was first made in \cite{BM}. In particular it was
shown, that because of saturation the multiplicity and transverse
energy density of gluons produced at central rapidity scales as
\begin{eqnarray}\label{mebm}
 {dN \over dy} & = & 2\ A xG_{nucleon}(x,Q^2_s), \nonumber \\
 {d E_\perp \over dy} & = & 2\ Q_s xG_{nucleon}(x,Q^2_s),
\end{eqnarray}
where $Q_s \sim \alpha A /R^2$ is a characteristic saturation
scale at which gluon emission and recombination equilibrate. We
see that the $A$ - counting in Eq.~(\ref{mebm}) is different from
the naively expected perturbative factor $A^2$, and is more akin
to the one in soft production models.

It is important to stress that the physical picture of gluon
production, and thus that of the initially produced gluonic
configuration, depends on the gauge used in the calculation. This
was explicitly demonstrated in \cite{MK99}, where a calculation of
the spectrum of gluons produced in p-A collisions was performed
both in covariant and light-cone gauge. It turned out that the
{\it origin} of A-dependent effects looks completely different in
these two gauges: in the covariant gauge it is a rescattering of
the produced gluon on the nucleons in the nucleus, and in the
light-cone gauge it is a nonlinear interaction of gluons in the
nuclear wavefunction. This explains a certain ambivalence with
which an intuitive reasoning explaining the basic features of
gluon production is formulated, see, e.g., \cite{AM00a}. Here it
will be convenient to follow the "light-cone gauge" way of
reasoning, in which the number of produced gluons can be expected
to be roughly proportional to the their "pre-existing" number in
the nuclear wavefunction \cite{AM00a}.

\subsubsection{Spectrum of produced gluons: analytical
results}\label{an}


The qualitative ideas described in the previous subsection were
further developed in \cite{K0102}, where the fully nonlinear
analytical  ansatz for the spectrum of gluons produced in the
collision of two identical nuclei was suggested. The corresponding
formula in \cite{K0102} can be written as a two-dimensional
integral integral in the transverse coordinate space. With
logarithmic accuracy and for parametrically small transverse
momenta $k^2_\perp < Q^2_s$ one can perform the two-dimensional
integration over coordinates and obtain the following impressively
simple expression for the gluon spectrum:
\begin{equation}\label{specaa}
 {d N^{AA} \over d^2b dy d^2k_\perp} \, = \,
 {C_F \over \alpha \pi^3}\ {Q^2_s \over k^2_\perp} \left [
 {\rm e}^{-k^2_\perp/2 Q^2_s} - {\rm e}^{-k^2_\perp/Q^2_s}
 \right ].
\end{equation}
From Eq.~(\ref{specaa}) there follows an important conclusion that
(up to possible logarithmic factors neglected in the process of
its derivation) the spectrum of gluons produced in nucleus-nucleus
collision is {\it finite} in the limit $k^2_\perp / Q^2_s \to 0$:
\begin{equation}
 {d N^{AA} \over d^2b dy d^2k_\perp}
  \to
 {1 \over \alpha} {C_F \over 2 \pi^3},
\end{equation}
thus ensuring an infrared-finite answer for quantities containing
integration over transverse momenta such as, e.g., inelastic
cross-section. This result is highly nontrivial. In the standard
minijet scenarios based on collinear factorization such infrared
finiteness can be ensured only by using brute force (an explicit
infrared cutoff for a strongly divergent spectrum  $\sim
1/k^4_\perp$). In p-A scattering, where nonlinear corrections
related to the single participant nucleus are summed, the gluon
spectrum still possesses a powerlike divergence at small momenta
($\sim 1/k^2_\perp$) \cite{KM98,K0102}. This shows that it is only
a combination of all nonlinear effects in both colliding nuclei
that ensures the infrared finiteness of the spectrum of produced
gluons and the infrared finiteness of physical cross-sections
computed from it.

The spectrum of Eq.~(\ref{specaa}) allows to make quantitative
estimates relating the physical quantities to the saturation
momentum more precise. In particular, the mean transverse momentum
of produced gluons reads
\begin{equation}
\langle k^2_\perp \rangle \, = \, {1 \over \ln 2}\ Q^2_s
\end{equation}
We see, that the numerical value $\langle k^2_\perp \rangle$ is
indeed very close to that of $Q^2_s$ supporting the intuitive
picture advocated in \cite{MV94} and \cite{BM,AM00a}. Performing
in Eq.~(\ref{specaa}) integration over $k_\perp$, we obtain an
expression for the gluon rapidity density in the transverse plane
\begin{equation}\label{specaay}
 {d N^{AA} \over d^2b dy} \, = \, {1 \over \alpha}
 {\ln 2\, C_F \over \pi^2} Q^2_s
\end{equation}
It is illuminating to compare Eq.~(\ref{specaay}) with the
expression for the density of gluons in the nuclear wavefunction
computed in the same cylindrical geometry
\begin{equation}\label{ndkov}
 {d N^{MW} \over d^2b dy} \, = \, {1 \over \alpha}
 {C_F \over 2 \pi^2} Q^2_s
\end{equation}
Comparing Eq.~(\ref{specaay}) with Eq.~(\ref{ndkov}) we see that
the density of {\it produced} gluons Eq.~(\ref{specaay}) is indeed
proportional to the density of gluons in the nuclear wavefunction
\begin{equation}
{d N^{AA} \over d^2b dy} \, = \, 2 \ln 2 \, {d N^{MW} \over d^2b
dy}
\end{equation}
with the proportionality coefficient $2 \ln2 \simeq 1.39$. From
Eq.~(\ref{specaay}) one can also express the rapidity density of
the produced gluons in terms of the nucleon structure function:
\begin{equation}\label{specy}
 {d N^{AA} \over dy} \, = \, \pi R_A^2 {1 \over \alpha}
 {\ln 2 C_F \over \pi^2} Q^2_s
 \, = \, 2 \ln2 (V_A \rho) x G(x,Q^2_s),
\end{equation}
where $V_A \sim A$ is a nuclear volume.


\subsubsection{Parton production and saturation: numerical
solution.}\label{num}

A natural extension of the philosophy of McLerran-Venugopalan
approach to high energy heavy ion physics is to study, at the same
quasiclassical level, the spectrum of gluons produced in collision
of two nuclei, where the gluon spectrum is determined by the mode
content of the gluon field created by two colliding nuclei
\cite{KMW95}. In the pioneering paper \cite{KMW95} the spectrum of
produced gluons was calculated to the leading order in
perturbation theory. The answer contained a characteristic strong
infrared divergence. Later these calculations were expanded in
\cite{KR97,MMR98,GM97}. The problem at hand amounts to solving the
Yang-Mills equations in the presence of external source current
\cite{KMW95}:
\begin{equation}\label{colcur}
 J^\mu \, = \, \delta^{\mu +} \rho_{(1)} \delta (x^-) +
 \delta^{\mu -} \rho_{(2)} \delta (x^+)
\end{equation}
corresponding to the two incident nuclei. Full analytical solution
of the problem seems impossible, so a dedicated program of its
numerical analysis was launched \cite{KV99}-\cite{TL03}.  Assuming
the boost invariance of the problem one deals with the problem of
numerically solving the equations of motion  in (2+1) classical
hamiltonian chromodynamics on the lattice.

The result depends on three parameters,- charge $g$, color charge
density $\mu$ and nuclear radius $R_A$, - through their
dimensionless combination $\xi = g^4 \pi R^2_A \mu^2$, so that for
rapidity density of multiplicity and transverse energy of
primordial glue one has
\begin{eqnarray}\label{numdis}
 {d E_\perp \over dy} & = & \mu\,\xi\,f_E(\xi) \nonumber \\
 {d N \over dy} & = & \xi\,f_N(\xi).
\end{eqnarray}
In the weak field limit (more exactly, for $\xi < 50$) all
quantities are strongly dependent on $\xi$ and thus on the
infrared cutoff. At $\xi \sim 100$ this dependence saturates.

One important issue to analyze is to check the (weak field)
regime, in which the perturbative result of \cite{KMW95} should be
valid. Most resent analysis shows, that the agreement can be
reached only at very small values of scaling parameter $\xi < 10$
\cite{TL03}. As to the spectrum of produced gluons, it has an
exponential "thermal-looking" one at small energies
\cite{KNV01,TL03}, but deviates from it at large energies
\cite{TL03}.

The most important issue addressed by the numerical computation
is, probably, that of a magnitude of occupation numbers of gluon
modes $f_g$. The classical description that lies at the heart of
the method is justified only at large $f_g \gg 1$. Although the
situation here does not seem finally settled, for parameter values
corresponding to RHIC energies, this condition is satisfied at
best marginally.


\subsubsection{Interpreting the RHIC data in Color Glass Condensate
terms.}\label{satexp}


With a wealth of experimental data coming from RHIC it is tempting
to test the ideas of Color Glass Condensate (saturation) physics
in the simplest setting. Let us assume, that from the moment the
Color Glass Condensate melts into physical glue, the produced
gluons do not subsequently reinteract and convert into final
hadrons without changing the kinematical characteristics of the
energy-momentum flux (soft hadronization hypothesis). Then, by
comparing with experimental data on charged multiplicity
\cite{PHOB0002,PHEN01,BRAH01,STAR01} or multiplicity per
participant \cite{PHOB0002} one can constrain parameters
specifying the initially produced gluon configuration.

Consider for example the cylindrical nuclei model discussed in
\cite{K0102}. Then using, for example, the charged multiplicty
density measured by PHOBOS in Au-Au collisions \cite{PHOB0002}
\begin{equation}
{d N_{ch}^{AuAu} \over d \eta} \, = \, 555 \pm 12(stat) \pm
35(syst)
\end{equation}
one obtains from Eq.~(\ref{specy}), taking $\alpha_s=0.3$ and $\pi
R_A^2=150$ fm,  the estimate for the saturation momentum $Q_s$:
$Q^2_s \approx 0.7 \,\, {\rm GeV}^2$.


The relation of the results of numerical lattice computations
described in paragraph \ref{num} to the experimental data could be
done analogously. For example, using the second of
Eqn.~(\ref{numdis}), one can determine (for given $g$ and $R_A$)
the value of $\mu$ and compute from the first of
Eqn.~(\ref{numdis}) the transverse energy density. Using
\cite{TL03} $dN/dy \approx 1000$, one gets (for $g=2$ and
$S_A=150$ fm$^2$)  $\mu=0.5$ GeV and $dE_\perp/dy=1.5\ {\rm GeV}\
dN/dy$.


A more elaborate way of estimating the characteristics of Color
Glass Condensate from experimental data was suggested in
\cite{KN01,KL01,KLN01}. The main novelty of this approach is to
use a density of participating nucleons in the formula determining
the saturation scale, so that
\begin{equation}
 Q^2_s(s_\perp,b_\perp) \, = \, {4 \pi^2 N_c \over N^2_c-1}\
 \alpha_s (Q^2_s)\ xG(x,Q^2_s)_{nucleon}\
 {\rho_{part} (s_\perp,b_\perp) \over 2}
\end{equation}
where $\rho_{part} (s_\perp,b_\perp)$ is a density of participant
nucleons as a function of the collision impact parameter $b_\perp$
and the coordinate in the transverse plane $s_\perp$. The
substitution of $\rho_{part}$ into equation determining a
saturation scale presents a highly nontrivial hypothesis on, in
fact, non-perturbative geometry present behind gluon production.
The resulting relation between multiplicity per participant and
saturation momentum reads \cite{KN01,KL01,KLN01,BMSS02,M02}:
\begin{equation}\label{parden}
\left \langle {2 \over N_{part}}\ {dN_{ch} \over dy} \right
\rangle \simeq {2 \over 3}\ c\ x G(x,\langle Q^2_s \rangle ),
\end{equation}
where $c$ is a proportionality coefficient between the gluon
spectrum and nuclear wavefunction discussed in paragraph \ref{an},
averaging in the left-hand side is over events having different
number of participants and $\langle Q^2_s \rangle$ denotes
averaging over the impact parameter. Experimentally $2/N_{part}
dN_{ch}/dy \simeq 3.8$, so that from Eq.~(\ref{parden}) one can
estimate the (average) saturation momentum $Q_s$. Without invoking
additional assumptions the typical value for $Q^2_s$ on gets from
Eq.~(\ref{parden}) is also not too big: $Q^2_s \simeq 0.5 - 0.7$
GeV$^2$.

\subsection{Interaction effects. On the way to thermalization?}

Up to now we have discussed only the properties of the initially
produced gluon system appearing immediately after the coherence of
the wavefunctions of incident nuclei is broken by the collision
and, as a consequence, entropy in the form of the physical (mainly
gluonic) fields is produced. Before the energy-momentum flux of
these fields converts into that of final hadrons hitting detectors
it could, however, be essentially transformed by interaction
effects. The question that has particularly shaped the high energy
heavy ion physics is whether the reinteraction of produced parton
matter could lead to its thermalization into quark-gluon plasma,
thus allowing to reproduce, in the laboratory, conditions that
existed in the Early Universe. In this paragraph we shall briefly
review the recent progress in describing the real time evolution
of an interacting (dense) gluon system.

Broadly speaking, one could classify reinteraction effects into
two categories.

First, if a strong {\it physical} gluon field is produced, it can
evolve (in real time) according to the nonlinear Yang-Mills
equations of motion. This regime is possible up until the
occupation numbers of the field modes become small. Schematically,
the occupation numbers $f$ should satisfy $1 < f < 1/\alpha_s$.
Such nonlinear evolution could, in principle, lead to all kind of
exciting scenarios typical for the nonlinear field dynamics - from
appearence of collective dynamical instabilities to chaotization
and formation of solitons.

Second, one could describe the reinteraction of produced physical
glue in terms borrowed from kinetic theory \footnote{There are
good grounds to believe that these two approaches are (at least
partially) complementary \cite{MS03}.}. This possibility has been
discussed, in relation to saturation physics in nuclear
collisions, in a number of recent publications
\cite{AM00b}-\cite{BMSS02}.

The simplest way to analyze gluon reinteraction effects is to use
a Boltzmann equation formalism at a binary scattering level
\cite{AM00b,BV00,SS01}. Calculation of equilibration time in this
approximation produce a parametrically big estimate $\tau_{eq}
\sim {\rm exp}(1/\sqrt{\alpha_s}) 1/Q_s$. The equilibration rate
is low because the momentum transfer in the system is not
effective: the transverse momenta exchanged in gluon interactions
are small - of order of the infrared cutoff (screening Debye
mass).

The main motivation for developing a saturation physics approach
is a very dense system of primordial gluons that is, presumably,
formed at the initial stage of high energy heavy ion collision. To
produce a more reliable description for gluon reinteraction and
their possible equilibration a kinetic approach that that is more
appropriate for (initially) dense systems is called for. Such
approach was developed in \cite{BMSS01,BMSS02}, where the kinetic
equation formalism taking into account inelastic processes in the
third order in gluon density was constructed and employed. The
resulting reinteraction scenario described in \cite{BMSS01,BMSS02}
is quite complex and involves several stages. The corresponding
proper time scales are $\tau \sim
(\tau_0,\alpha_s^{-3/2}\tau_0,\alpha_s^{-5/2}\tau_0,\alpha_s^{-13/5}\tau_0))$,
where $\tau_0 = 1/Q_s$.

First, at $\tau \sim \tau_0$, physical glue is freed from the
nuclear wavefunctions. This is a dense system of (semi)hard gluons
having transverse momenta of order of $Q_s$ and occupation number
of order $1/\alpha_s$. The system expands, and at $\tau \sim
\alpha_s^{-3/2}\tau_0$ the occupation numbers of semihard
primordial gluons become small, so that a standard description in
terms of Boltzmann equation can be applied.

In the time interval $\alpha_s^{-3/2}\tau_0 < \tau <
\alpha_s^{-5/2}\tau_0$ inelastic interactions of semihard gluons
produce soft gluons with momenta $k_\perp \sim \alpha_s^{1/2}$. At
the end of this interval the densities of hard and soft gluon
components equalizes.

At $\tau > \alpha_s^{-5/2}\tau_0$ the soft gluon subsystem
thermalizes. Its temperature subsequently undergoes a linear
increase through the energy loss of the remaining semihard modes
in the hot soft gluon medium until it reaches, at $\tau \sim
\alpha_s^{-13/5}\tau_0$, its maximal value $T \sim \alpha_s^{2/5}
Q_s$. At this timescale semihard glue disappears and the gluon
system is fully equilibrated.

Detailed discussion of RHIC data in the context of the
above-described scenario can be found in \cite{BMSS02}.

Let us note, that the validity of the scenario described in
\cite{BMSS01,BMSS02} is based on some quite restrictive
assumptions. For example, for equilibration time
$\alpha_s^{-13/5}\tau_0$ to be less than the "binary" one ${\rm
exp}(1/\sqrt{\alpha_s})$, the coupling constant should be really
small, $\alpha_s < 0.004$. Also - especially at RHIC energies,
when $Q_s \sim 1\,{\rm GeV}^2$ and realistic values of the
coupling constant $\alpha_s \sim 0.3$, the transverse momenta of
the soft gluons produced to the second stage $k_\perp \sim
\alpha_s^{1/2} \tau_0$ are in fact of order $\Lambda_{QCD}$, so
that to describe the evolution of the soft gluon subsystem
perturbative methods could turn out to be insufficient.



\section{Conclusion}

In the present review we have discussed some aspects of an
exciting and rapidly developing field of nonlinear QCD physics in
ultrarelativistic heavy ion collisions. Research in this field
deals both with fundamental theoretical issues, such as unitarity
of strong interactions at high energies, and with the challenge of
describing experimental data coming, at present, from RHIC and
expected exciting physics of forthcoming experiments at LHC.

\begin{center}
{\bf Acknowledgements}
\end{center}

I am indebted to I.M.~Dremin, E.~Iancu, A.~Kovner and L.~McLerran
for reading the manuscript and useful comments and suggestions.



\end{document}